\documentclass[%
 reprint,
 amsmath,amssymb,
 aps,
]{revtex4-2}

\usepackage{graphicx}
\usepackage{dcolumn}
\usepackage{bm}
\usepackage{xcolor}
\usepackage{svg}
\usepackage{times}

\begin{document}

\preprint{APS/123-QED}

\title{Competing Relaxation Channels in Continuously Polydisperse Fluids: A Mode-Coupling Study}

\author{Corentin C.L. Laudicina}%
 \affiliation{Soft Matter and Biological Physics, Department of Applied Physics, Eindhoven University of Technology,
P.O. Box 513, 5600 MB Eindhoven, Netherlands}
\author{Ilian Pihlajamaa}
 \affiliation{Soft Matter and Biological Physics, Department of Applied Physics, Eindhoven University of Technology,
P.O. Box 513, 5600 MB Eindhoven, Netherlands}
\author{Liesbeth M.C. Janssen}%
\email[Contact: ]{l.m.c.janssen@tue.nl}
 \affiliation{Soft Matter and Biological Physics, Department of Applied Physics, Eindhoven University of Technology,
P.O. Box 513, 5600 MB Eindhoven, Netherlands}

\date{\today}

\begin{abstract}
We perform a systematic analysis of continuously polydisperse hard-sphere fluids as a function of the degree of polydispersity within the framework of the Mode-Coupling Theory of the glass transition (MCT). Our results show that a high degree of polydispersity tends to stabilize the liquid phase against vitrification, the magnitude of which depends on the shape of the polydispersity distribution. Furthermore, we report on a separation between the localization lengths of the smallest and largest particles. A diameter-resolved analysis of the intermediate scattering functions reveals that this separation significantly stretches the relaxation patterns, which we quantitatively study by an analysis of the dynamical exponents predicted by the theory. Our observations have strong implications for our understanding of the nature of dynamical heterogeneities and localization lengths in continuously polydisperse systems with hard-core interactions. These results suggest that the dynamics of the smallest particles is of central importance to understand structural relaxation of such systems, already in the mildly supercooled regime where MCT is usually applicable. 
\end{abstract}

\maketitle

\section{Introduction}

Size polydispersity, i.e.\ a heterogeneity in particle sizes, is often inevitable and even necessary to produce stable supercooled liquids. Indeed, without such dispersity, most liquids and colloidal suspensions soon crystallize upon supercooling or compression \cite{kawasaki2011structural, berthier2023modern}. In fact, size-polydisperse systems form the bulk of the modern literature on computational studies of the glass transition, in particular with the advent of enhanced sampling procedures such as swap Monte Carlo, which require extremely high degrees of size polydispersity for high efficiency \cite{ninarello2017models, berthier2023modern}. While size polydispersity has been shown to stabilize the liquid phase in the supercooled regime, for both simple colloidal systems \cite{pusey2009hard, leocmach2013importance, ninarello2017models, berthier2023modern} as well as for polymeric ones \cite{li2016influence}, the introduction of different particle species also inherently complicates the structural and dynamical aspects of glass-forming systems \cite{schope2007effect, coslovich2018local, baranau2020relaxation}. In particular, the behavior of the smallest particles appears to be of crucial importance for the system's global structural relaxation \cite{zaccarelli2015polydispersity, pihlajamaa2023polydispersity}. 

From a theoretical perspective, the Mode-Coupling Theory of the glass transition (MCT) is the only microscopic framework capable of making quantitative predictions on the dynamics of dense, supercooled liquids \cite{leutheusser1984dynamical,bengtzelius1984dynamics, gotze2009complex} from purely structural considerations (i.e.\ the liquid's pair correlation function). The theory provides a set of equations of motion for several dynamical observables: the coherent and incoherent intermediate scattering functions, as well as the mean-square displacement if appropriate limits are taken \cite{gotze2009complex}. In the case of simple glass formers, MCT predicts a phase transition to an ideal glass state which can be rigorously studied using techniques from the bifurcation of iterated maps \cite{gotze1996bifurcations}. Beyond this, MCT's principal successes lie in the prediction of accurate form factors (e.g.\ Debye-Waller and Lamb-M\"ossbauer (LM) factors) as well as the hallmarked two-step, and stretched exponential relaxation of intermediate scattering functions. Furthermore, the theory predicts non-trivial associated scaling laws and dynamical exponents (and algebraic relations between them) which many liquids in the supercooled regime have been found to obey both in computer simulations and experimentally \cite{li1992testing, kob1995testing, gotze1999recent,sciortino2001debye}. Moreover, the theory can describe more intricate dynamics such as re-entrant behavior in e.g.\ mildly asymmetric binary mixtures \cite{gotze2003effect}, systems with short-range attractions \cite{dawson2000higher,pham2002multiple} or systems in strong geometric confinement \cite{mandal2017nonergodicity}, as well as more exotic phase transitions, in particular glass-glass transitions \cite{fabbian1999ideal, dawson2000higher, voigtmann2011multiple} and partially arrested states \cite{thakur1991glass, voigtmann2011multiple}. Recent work has also demonstrated that the theory is capable of making predictions on dynamical heterogeneities, something which was initially deemed impossible due to the inherent absence of many-body correlations in the theory \cite{gotze2009complex}. Indeed, a careful treatment reveals that MCT predicts a diverging (dynamical) correlation length associated with its ideal glass transition \cite{biroli2006inhomogeneous}.

Within the context of MCT, polydispersity-induced effects have been widely studied in binary and ternary mixtures, including for large size ratios (i.e.\ high polydispersity degrees) \cite{flenner2005relaxation, voigtmann2003mode}. Furthermore, Weysser \textit{et al.}\ \cite{weysser2010structural} have tested MCT for a system of mildly continuously polydisperse ($\sim 5\%$) quasi-hard spheres where the diameters are drawn from a uniform distribution. By performing an appropriate multi-component analysis of the mixture, they reported good agreement for the collective and single-particle observables predicted by the theory, except at low wave-numbers for the latter (a typical shortcoming of the theory, especially for single-particle observables). Their work has clearly highlighted the importance of treating polydispersity accurately by comparing predictions from effective monodisperse MCT and multicomponent MCT for up to five components. Beyond this, little is known on how the degree of size polydispersity quantitatively affects MCT's predictions in continuously polydisperse mixtures.

This work aims to extend these prior studies by providing a systematic and quantitative study of the glassy dynamics predicted by MCT as the degree of polydispersity of continuously polydisperse mixtures is varied. The focus of this work is on MCT predictions solely, without resorting to explicit comparisons with experimental or simulation results. There are various reasons for this choice; firstly, this removes the need for computationally intensive simulations since the MCT equations of motion can be solved within reasonable computational time, at least up to ten different particle components using the latest open-source code implementation of MCT \cite{MCT_solver}. Furthermore, there are already many experimental and computational studies that have demonstrated that continuously size polydisperse mixtures display non-trivial dynamical features \cite{abraham2008energy, zaccarelli2015polydispersity, pihlajamaa2023polydispersity} which are by-products of the continuous aspect of polydispersity. Secondly, and perhaps more importantly, it is now universally accepted that MCT does not account for the experimentally observed dynamical crossover, which is a widely seen phenomenon that signals a change of dynamics upon entering the more deeply supercooled regime \cite{mallamace2010transport, berthier2011theoretical}.
 The physical mechanisms that drive the crossover are known as activated events, and are believed to be associated with increasingly cooperative relaxation \cite{vogel2004spatially,bergroth2005examination, candelier2010spatiotemporal, guiselin2022microscopic,scalliet2022thirty}. This in fact forms one of the main criticisms against the theory, and in practice, limits MCT's range of applicability to only the mildly supercooled regime. Here we turn this deficiency into an advantage, as it allows us to probe how the degree of polydispersity affects one well-known, and dominant relaxation mechanism in the mildly supercooled regime in a way that is completely independent from the other mechanisms that are beyond the scope of the theory.  
This enables us to disentangle potential polydispersity-induced effects which have remained unstudied in recent simulation works on very deeply supercooled liquids \cite{guiselin2022microscopic, scalliet2022thirty}, yet are present already at mildly supercooled states where MCT would be applicable.

\section{Theory}
\label{sec:methods}
We consider an equilibrium, underdamped mixture of $N$ particles which can be divided into $n$ distinct components, each made up of $N_{\sigma}$ particles of type $\sigma$, where $\sigma=1,2,\ldots,n$. We define the incoherent intermediate scattering functions (ISF) per species at wave-vector $\mathbf{k}$ as $F^{(s)}_{\sigma}(k, t) \equiv N_{\sigma}^{-1}\langle\sum_{j=1}^{N_{\sigma}} \exp[i \mathbf{k}\cdot(\mathbf{r}_j^{\sigma}(0)-\mathbf{r}_j^{\sigma}(t)] \rangle$, where $\mathbf{r}_i^{\sigma}(t)$ denotes the position of particle $i$ of species $\sigma$ at time $t$. We note that the ISF only depends on the wave-number $k =|\mathbf{k}|$ since we consider a translationally and rotationally invariant system. The angular brackets denote an ensemble average. Using standard projection operator methods, the incoherent ISF for species $\sigma$ is found to satisfy the following integro-differential equation
    \begin{equation}
    \begin{split}
    \ddot{F}^{(s)}_{\sigma}(k, t)&+\Omega_{\sigma}^{(s)}(k)^2 F_{\sigma}^{(s)}(k, t) \\
    &+\int_0^t d \tau K_{\sigma}^{(s)}(k, t-\tau) \dot{F}_{\sigma}^{(s)}(k, \tau)=0
    \end{split}
    \label{eq:tagged_multicomponent_MCT}
    \end{equation}
with $\Omega^{(s)}_{\sigma}(k)^2 \equiv k^2k_BT/m_{\sigma}$, where $m_{\sigma}$ is the mass of particle type $\sigma$, $k_B$ Boltzmann's constant and $T$ denotes the temperature. The MCT memory kernel $K^{(s)}_{\sigma}(k, t)$ reads
    \begin{equation}
    \begin{split}
        K^{(s)}_{\sigma}(k, t)=\frac{k_B T \rho}{k^3 m_{\sigma}} \int_{\mathbb{R}^3} &\frac{d\mathbf{q}}{(2\pi)^3}(\mathbf{k} \cdot \mathbf{q})^2 c_{\sigma \gamma}(q) c_{\sigma \lambda}(q) \\
        &\times F_{\gamma \lambda}(q, t) F^{(s)}_{\sigma}(|\mathbf{k}-\mathbf{q}|, t),
    \end{split}
    \end{equation}
where $\rho$ is the bulk density, and $c_{\alpha\beta}(k)$ is the partial two-body direct correlation function, which is related to the partial static structure factor $S_{\alpha\beta}(k)$ by $c_{\alpha\beta}(k) \equiv \rho^{-1}[\delta_{\alpha\beta}/x_{\alpha} - (\mathbf{S}^{-1}(k))_{\alpha\beta}]$ \cite{hansen2013theory}. Throughout this work, Greek indices refer to particle species labels. The superscript $(s)$ refers to a `self' (tagged) quantity. Furthermore, Einstein summation convention is used throughout the manuscript. 

Within MCT, The incoherent ISF depends on the coherent one, which is defined as

    \begin{equation}
        F_{\alpha \beta}(k, t) \equiv N^{-1}\left\langle \sum_{i=1}^{N_{\alpha}}e^{-i\mathbf{k}\cdot\mathbf{r}^{\alpha}_i(0)}\sum_{j=1}^{N_{\beta}}e^{i\mathbf{k}\cdot\mathbf{r}_j^{\beta}(t)}\right\rangle.
    \end{equation}
The coherent ISF satisfies the following equation of motion: 

    \begin{equation}
    \begin{split}
        \ddot{F}_{\alpha\beta}(k,t) &+ \Omega_{\alpha\gamma}(k)^2F_{\gamma\beta}(k,t) \\
        &+ \int_0^t d\tau K_{\alpha\gamma}(k,t-\tau)\dot{F}_{\gamma\beta}(\tau) = 0,
    \end{split}
    \label{eq:collective_multicomponent_MCT}
    \end{equation}
in which $\Omega_{\alpha\beta}(k)^2 \equiv k^2k_BTx_{\alpha}/m_{\alpha} \cdot (\mathbf{S}^{-1}(k))_{\alpha\beta}$ and where the collective memory kernel $K_{\alpha\beta}(k,t)$ is approximated by a bi-linear functional of the coherent ISF. This is the celebrated mode-coupling approximation, which reads

    \begin{equation}
    \begin{split}
        K_{\alpha\beta}(k,t) = \frac{k_BT\rho}{2m_{\alpha}x_{\beta}}\int_{\mathbb{R}^3} \frac{d\mathbf{q}}{(2\pi)^3} &V_{\alpha\gamma\eta}(\mathbf{k}, \mathbf{q})V_{\beta\kappa\epsilon}(\mathbf{k}, \mathbf{q}) \\
        &\times F_{\gamma\eta}(q,t)F_{\kappa\epsilon}(|\mathbf{k}-\mathbf{q}|, t).
    \end{split} \label{eq:Kmct}
    \end{equation}
We denote by $x_{\sigma}\equiv N_{\sigma}/N$ the fraction of particle species $\sigma$. The coupling constants $V_{\alpha\beta\gamma}(\mathbf{k},\mathbf{q})$ can be expressed in terms of the partial two-body direct correlation function \cite{bengtzelius1984dynamics, gotze2003effect, gotze2009complex}. Overall MCT thus culminates in a closed set of dynamical equations, which can be solved self-consistently once the system-specific static inputs (in the form of the static structure factors, bulk density, temperature, and particle mass) are known. 

Since our goal is to investigate how the degree of size polydispersity in continuously polydisperse mixtures modifies the theory's principal predictions \cite{franosch1997asymptotic, gotze2009complex}, let us briefly recall them. There exists a critical packing fraction $\varphi_c$ beyond which an `ideal glass' forms (i.e.\ the point where the density correlation functions no longer decay to zero at any finite time). This marks an ergodicity breaking point. When approached from the liquid side, the transition point is accompanied by a strict divergence of the structural relaxation time $\tau_{\alpha} \sim |\varphi-\varphi_c|^{-\gamma}$ with a critical exponent $\gamma$. The structural relaxation time $\tau_{\alpha}$ is typically defined as the point where the correlation functions have decayed to a small threshold value (e.g.\ $1/e$ or 0.1), although more formal definitions also exist \cite{franosch1997asymptotic}.  

Close to $\varphi_c$, both the coherent and incoherent ISF develop a distinct two-step relaxation pattern with a long-lived plateau at intermediate times. Around this plateau, the ISFs admit the following asymptotic expansions to leading order:

\begin{equation}
F_{\alpha\beta}(k,t) =
    \begin{cases}
        f_{\alpha\beta}(k) + K_{\alpha\beta}^{(+)}(k)(t/t_0)^{-a} +\mathcal{O}((t/t_0)^{-2a})\\
        f_{\alpha\beta}(k) - K_{\alpha\beta}^{(-)}(k)(t/t_0)^{b} +\mathcal{O}((t/t_0)^{2b})
    \end{cases}
\end{equation}
and
\begin{equation}
F^{(s)}_{\sigma}(k,t) =
    \begin{cases}
        f^{(s)}_{\sigma}(k) + K_{\sigma}^{(+)}(k)(t/t_0)^{-a}+\mathcal{O}((t/t_0)^{-2a}) \\
        f^{(s)}_{\sigma}(k) - K_{\sigma}^{(-)}(k)(t/t_0)^{b}+\mathcal{O}((t/t_0)^{2b}),
    \end{cases}
\end{equation}
where $f_{\alpha\beta}(k)$ denotes the plateau height of the partial coherent ISF [also known as the partial Debye-Waller (DW) factor], and $f^{(s)}_{\sigma}(k)$ that of the incoherent contributions to the ISF [also known as the Lamb-M\"{o}ssbauer (LM) factor] respectively. One should consider the $(+)$ solutions on the approach towards the plateau, and the $(-)$ solutions upon departure from it. The quantity $t_0$ is some reference timescale which determines the range of validity of the asymptotic expansion, usually defined as the point at which the coherent (incoherent) ISF equals the value of the associated DW (LM) factor at the critical point \cite{franosch1997asymptotic}. The factorization theorem \cite{gotze2009complex} indicates that every term in the asymptotic series separates into a purely wave-number-dependent contribution $K$ (also known as the critical amplitude), and a solely time-dependent function which, to leading order behaves as a power-law with exponents $a,\ b$ depending on whether one is above or below the critical plateau. The exponent $b$ is commonly known as the von-Schweidler exponent \cite{vonSchweidler1907studien}. Furthermore, the three dynamical exponents are related by the following two relations
    \begin{equation}
        \gamma = \frac{1}{2a} + \frac{1}{2b},\ \  \text{ and }\ \  \frac{\Gamma(1-a)^2}{\Gamma(1-2a)} = \frac{\Gamma(1+b)^2}{\Gamma(1+2b)},
    \label{eq:MCT_exponent_relations}
    \end{equation}
where $\Gamma(x)$ denotes the gamma function. 
We note that the exponents are system dependent, but for a given system they are universal for the correlation functions discussed above. Since all species share the same exponents, the species averaged quantity $\overline{F^{(s)}}(k,t) \equiv x_{\sigma}F^{(s)}_{\sigma}(k,t)$ also admits the same asymptotic expansions, albeit with different critical amplitudes.  

In addition to accounting for the highly non-trivial scaling laws discussed above, MCT also provides a physically intuitive picture for glassy dynamics in terms of the cage effect. Briefly, this feedback effect stems from the self-consistent, non-linear form of the memory kernel of Eq.~\eqref{eq:Kmct}, which renders MCT incredibly sensitive to subtle changes in the microstructure of the liquid (as quantified by the static structure factor). In particular at the wave-vector corresponding to the first peak of the static structure factor, which typically changes the most upon supercooling, the relaxation dynamics is predicted to slow down significantly, which in turn also drives the slowdown of all surrounding wave-vectors via the self-consistent mode-coupling. Since this process is initiated at the length scale of the first solvation shell, the microscopic origin of the macroscopic slowdown is attributed to caging, whereby particles become trapped in local cages formed by their nearest neighbors. This ultimately culminates in the collective freezing of density fluctuations at the ideal glass transition. 

 \section{Numerical methods}

In the remainder of this work, we examine in detail how the degree of size polydispersity affects the dynamical behavior of a three-dimensional hard-sphere fluid near the ideal MCT transition. We study results for three different continuous distributions $P(D)$ of particle diameters $D$: 

    \begin{itemize}
        \item A uniform distribution $P_{\text{uni.}}(D)$ defined as
            \begin{equation}
        P_{\text{uni.}}(D) = \frac{1}{(D_{\text{max}} - D_{\text{min}})},\ D \in [D_{\text{max}},D_{\text{min}}]
    \end{equation}
where $D_{\text{max}}$ and $\ D_{\text{min}}$ denote the maximal and minimal diameters of the distribution respectively.
        \item A Gaussian distribution $P_{\text{Gauss}}(D)$ defined as 
        \begin{equation}
        P_{\text{Gauss}}(D) = \frac{1}{\sqrt{2\pi\Delta^2}}e^{-(D-\overline{D})^2/\Delta^2}
    \end{equation}
with mean $\overline{D}$ and standard-deviation $\Delta$.
\item An inverse cubic distribution $P_{\text{inv.}}(D)$ defined as
    \begin{equation}
        P_{\text{inv.}}(D) = \frac{A}{D^3}, \ D \in [D_{\text{max}},\ D_{\text{min}}]
    \end{equation}
where $D_{\text{max}}$ and $\ D_{\text{min}}$ denote the maximal and minimal diameters of the distribution respectively, and the normalization reads $A = 1/2 + 1/(\sigma-1)$ where $\sigma\equiv D_{\text{max}}/D_{\text{min}}$ is the size ratio of the largest to the smallest particle \cite{ninarello2017models}.
    \end{itemize}
     
The three distributions are shown on the right-most panels of Fig.~\ref{fig:phase_diagram}. 
We focus on different probability distributions as recent work has shown that not only the degree of polydispersity, but also the shape of the distribution matters near dynamical arrest \cite{zaccarelli2015polydispersity}. We note that the inverse cubic distribution has recently become very popular in computational studies of deeply supercooled liquids, owing to its exceptional equilibration efficiency when combined with the enhanced sampling swap Monte Carlo method  \cite{ninarello2017models}. 

To numerically solve Eqs.~\eqref{eq:tagged_multicomponent_MCT}-\eqref{eq:collective_multicomponent_MCT}, we approximate the size distributions by discretized versions. The particles' effective diameters are determined from a quantization of the probability distributions into $n=10$ quantiles (i.e.\ the distribution of particle diameters is sampled uniformly and thus a randomly selected particle has an equal probability of belonging to any particular bin). The effective diameter of particles of each bin is then set equal to the center of mass of the given bin. We show in Appendix \ref{app:quantisation} that an effective 10-component description is sufficient for the regimes of polydispersity investigated in this work. We also rescale the quantized diameter list such that it has unit mean. The width of the distributions (and thus the degree of size polydispersity) can be tuned by varying the range of the allowed diameters ($D_{\text{min.}},\ D_{\text{max.}}$ for the Uniform and Inverse Cubic distributions) or the half-width $\Delta$ for the Gaussian distribution. In order to allow for a fair comparison between different distributions, we characterize the degree of polydispersity by a polydispersity index denoted $\delta$, defined as the standard deviation of the (quantized) diameter list. Since the mean is fixed, the value of $\delta$ implicitly sets the upper and lower bounds of the particle diameters for a given distribution $P(D)$.

To permit a completely fit-parameter free, first-principles study, we use analytic partial static structure factors for hard spheres obtained from the multicomponent version of the Percus-Yevick (PY) approximation \cite{baxter1970ornstein}. These serve as our structural input to the MCT equations at a given density. We note that the multicomponent PY approximation has been used in polydisperse MCT studies in the past, leading to convincing predictions compared to simulation results \cite{voigtmann2004tagged}. Furthermore Frenkel \textit{et al.}\ \cite{frenkel1986structure} have demonstrated that the PY approximation was capable of accurately capturing the structure of (highly) continously polydisperse hard-sphere mixtures of log-normally distributed diameters. \\

The equations of motion \eqref{eq:tagged_multicomponent_MCT}-\eqref{eq:collective_multicomponent_MCT} are solved for the effective 10-component mixture over a logarithmically coarse-grained time grid \cite{fuchs1991comments,flenner2005relaxation}, which has been recently been implemented in an open-source solver for integro-differential equations of the mode-coupling type \cite{MCT_solver}. The wave-vector integrals were performed (in spherical coordinates) over a grid of 100 equidistant points between $k_{\text{min}} = 0.2$ and $k_{\text{max}} = 39.8$. We set the mass of all particles as well as the thermal energy to unity: $m_{\sigma}=1$ $\forall \ \sigma$ and $k_BT=1$. More details on the numerical routines, in particular with respect to numerical stability, can be found in Appendix \ref{app:numerical_details}. 

\section{State Diagram}

\begin{figure}[htbp]
    \centering
    \includegraphics[width=\columnwidth]{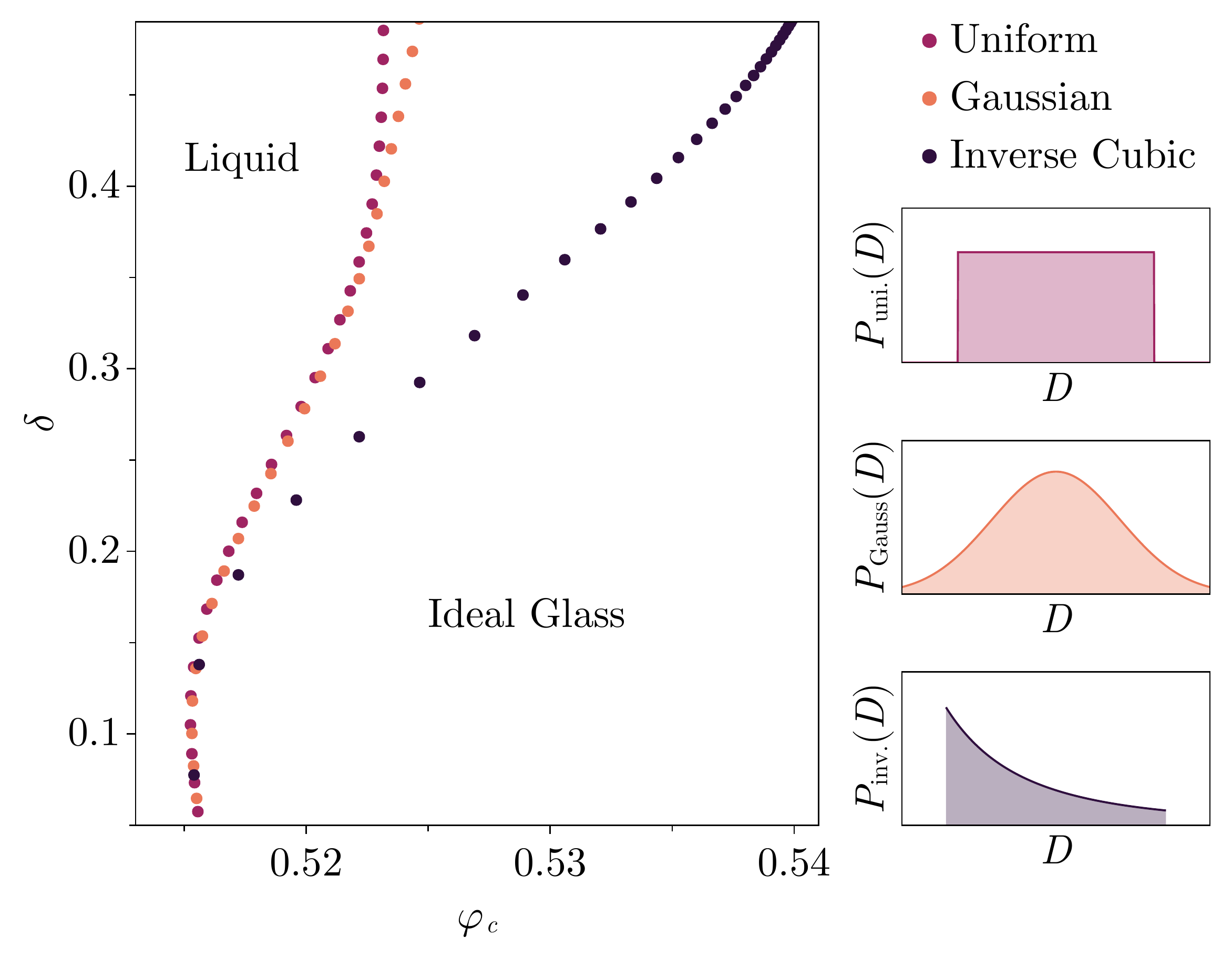}
    \caption{\textit{MCT state diagram in the polydispersity-packing fraction plane for polydisperse mixtures of hard-sphere liquids whose diameters are sampled from the three distributions described in the main text. The panels on the right-hand side illustrate the distributions of particle sizes $P(D)$.}}
    \label{fig:phase_diagram}
\end{figure}

We first study the location of the predicted glass transition for continuously polydisperse systems as a function of the degree of polydispersity. In order to determine the  state diagram, we solve Eq.~\eqref{eq:collective_multicomponent_MCT} for the long time limit ($t\rightarrow \infty$) by using a Picard iteration procedure and performing a bisection search for the critical point $\varphi_c$. In Fig.~\ref{fig:phase_diagram} we show the state diagram as determined by MCT for the three different distributions defined above. At low polydispersities, we find re-entrant behavior for all three cases, which has also been observed in the past in mildly size asymmetric binary hard-spheres in the context of MCT \cite{gotze1991liquids, gotze2003effect, foffi2003mixing, sciortino2005glassy}. As the polydispersity index is increased further, we see that the critical points are shifted to higher packing fractions for all distributions considered. This indicates that, according to MCT, a higher degree of size polydispersity stabilizes the liquid phase. This trend has also been observed in recent simulation studies of continuously polydisperse systems \cite{zaccarelli2015polydispersity}, and in studies of strongly size-asymmetric binary mixtures, both in simulation  \cite{foffi2003mixing} and within the context of MCT \cite{gotze2003effect}.
Physically, the shift of $\varphi_c$ toward higher values with increasing $\delta$ can be attributed to entropically induced effective attractions between particles with large size differences \cite{asakura1958interaction}, which are known to stabilize the liquid phase \cite{bergenholtz1999nonergodicity, dawson2000higher}.

For all three distributions studied here, considering a line at fixed packing fraction $\varphi$ could therefore lead to non-monotonic dependence of the relaxation time. This is particularly true near packing fractions close to that of the single component MCT transition $\varphi = 0.519$. For instance, at this point in the low to moderate polydispersity regime ($\delta < 0.15$), one initially expects an increase of relaxation time as the degree of polydispersity increases, followed by a glass region which eventually re-enters the liquid regime at high degrees of polydispersities, for $\delta\geq0.2$. In the latter regime we observe that the dynamics become faster as the degree of polydispersity is increased. This last observation is in line with prior studies on polydisperse Lennard-Jones fluids, where increasing the degree of polydispersity also leads to a speed-up of the dynamics \cite{abraham2008energy}.

It can be seen that the uniform and Gaussian distribution yield nearly identical critical points, except at very high polydispersities. The stabilization of the liquid phase is largest for the inverse cubic distribution, for which the critical packing fraction can be shifted beyond $\varphi_c = 0.540$. 
Note that this particular distribution also permits supreme equilibration efficiency in the supercooled liquid phase with swap Monte Carlo \cite{ninarello2017models}; whether these two observations bear the same physical origin deserves further investigation.  
We also recall that the nature of the inverse cubic distribution is such that there is an abundance of small particles compared to the large ones. Our findings therefore indicate, on purely theoretical grounds, that not only the relative size of the particles is crucial for the system's stability against vitrification but also the number of particles of each size. This also explains the deviation of the Gaussian distribution from the uniform one for $\delta>0.4$, as the former has more small particle outliers at fixed $\delta$, which would fluidize the system. 

In the limit of very high polydispersity indices $\delta$, we find that for the uniform distribution, the critical point saturates to a constant value.  This suggests that, at least at the level of MCT, there is a `maximal' polydispersity index beyond which the long-time dynamics are no longer affected. This behavior has also been observed in highly-asymmetric binary mixtures in the past \cite{gotze2003effect}. The saturation at large $\delta$ is more subtle in the case of the Gaussian and inverse cubic distributions, but we anticipate it should also be reached at sufficiently high polydispersities ($\delta>0.5$). We note however that at such high degrees of polydispersity, we cannot rule out the existence of exotic partially arrested states, as is predicted by MCT for binary mixtures where the size ratio of the smallest to the largest particles is below $D_{\text{min}}/D_{\text{max}} \leq 0.35$ \cite{voigtmann2011multiple}. In the present study, the size ratio $D_{\text{min}}/D_{\text{max}} \leq 0.25$ and $0.20$ for the uniform and Gaussian distributions of diameters respectively, if $\delta \geq  0.40$. It is therefore not unthinkable that a partially arrested glass exists in this portion of parameter space. The size ratio for the inverse cubic distribution remains smaller for equivalent values of $\delta$, although the same scenario is anticipated for higher degrees of polydispersity. The study of these more exotic glass states in continuously polydisperse mixtures is however beyond the scope of this work and is left for future study.

Lastly, we have verified that the same MCT state diagram is found for a finer wave-vector grid (see Appendix Fig.~\ref{fig:phase_diagram_fine_grid}), as well as for other quantizations for the particle diameters (fewer bins), which indicates that all the trends observed in Fig.~\ref{fig:phase_diagram} are genuine and not an artifact of our numerical routines.

In the remainder of this work, we focus on the dynamics of polydisperse systems where the particle diameters follow a uniform or inverse cubic distribution. This choice is motivated by the similarity between the MCT predictions for the Gaussian distribution and the uniform one. Furthermore, all the results that follow have been evaluated at the wave-number $k$ corresponding to the peak of the species-averaged static structure factor $\overline{S}(k) \equiv \sum_{\alpha\beta} S_{\alpha\beta}(k)$. This peak position is weakly polydispersity dependent near the critical point, as shown in Fig.~\ref{fig:struct_fact} (see Appendix \ref{app:structure}).

\section{Dynamical Features of Structural Relaxation} \label{dynamics_section}

\subsection{Incoherent Intermediate Scattering Functions and Form Factors}

\begin{figure*}
\includegraphics[width=\textwidth]{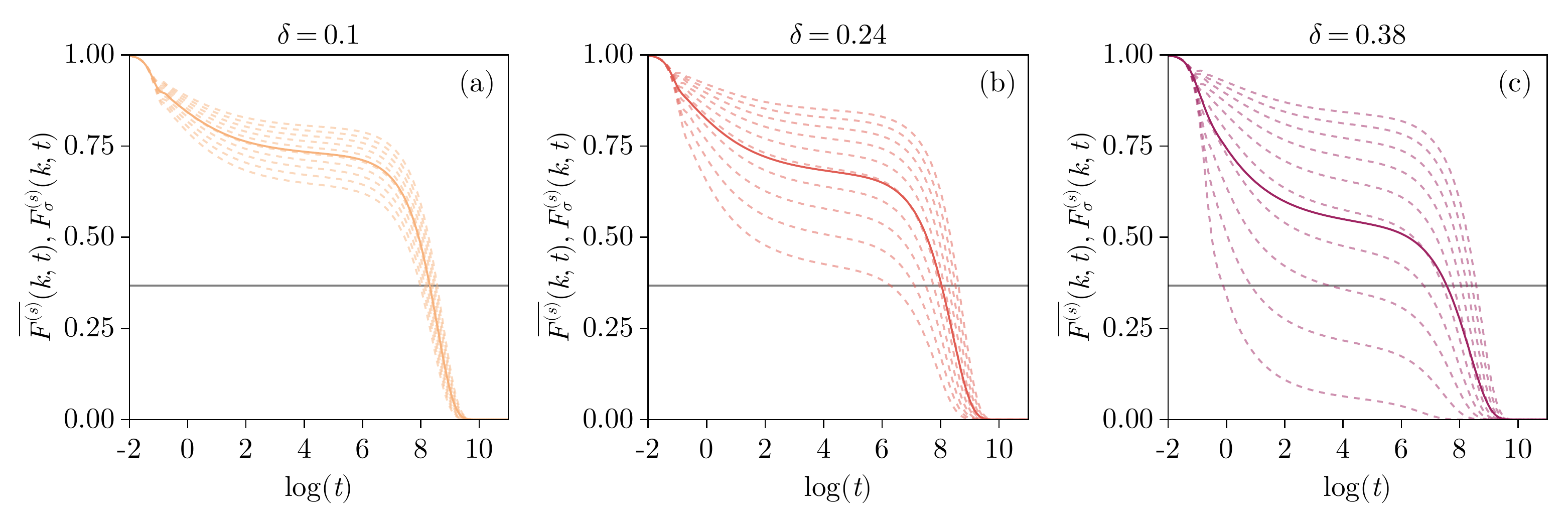}
\caption{\textit{Species-averaged (full line) and partial (dashed lines) incoherent intermediate scattering functions (denoted $\overline{F}_s(k,t)$ and $F_{\sigma}^{(s)}(k,t)$ respectively). The curves have been obtained by numerically solving for Eq.~\eqref{eq:tagged_multicomponent_MCT} for polydispersity degree $\delta = 0.1$} (a), 0.24 (b) and 0.38 (c) \textit{for a uniform distribution of diameters. The corresponding packing fractions are $\varphi = 0.515$} (a), 0.518 (b) \textit{and} 0.522 (c). \textit{For all three panels, the lowest dashed curve corresponds to the smallest particle diameter considered and the highest to the largest one (with a monotonic increase in between).}}
\label{fig:Fs_avg}
\end{figure*}

We now move on to study the dynamical features of structural relaxation in polydisperse mixtures. In order to isolate the effects of polydispersity, we compare systems at a fixed relaxation time $\overline{\tau}_{\alpha} = 10^{10}$ [a.u.], where $\overline{\tau}_{\alpha}$ is defined as the point where the diameter-averaged incoherent ISF has decayed to some small value $\varepsilon$, i.e.\ $\overline{F^{(s)}}(k,\overline{\tau}_{\alpha}) = \varepsilon$ with $\varepsilon = 10^{-5}$. We choose $\overline{\tau}_{\alpha} = 10^{10}$ [a.u.] to probe the dynamics fairly close to the MCT critical point, and $\varepsilon = 10^{-5}$ to ensure that all particle species have fully relaxed. One could also perform a study at fixed packing fraction $\varphi_c$ instead of fixed relaxation time $\overline{\tau}_{\alpha}$, while varying the polydispersity, but this would render the comparisons more difficult as $\overline{\tau}_{\alpha}$ is highly polydispersity dependent. 

We show in Fig.~\ref{fig:Fs_avg} the species-averaged (full line) and the partial incoherent ISF (dashed lines), for three different levels of polydispersity $\delta = 0.1,\ 0.24,\ 0.38$ for a uniform size distribution. In particular, looking at the partial incoherent ISF, we see that for $\delta = 0.24$ there is a decade of difference in the relaxation times for the smallest (lowest dashed line) and the largest particles (highest dashed line), and nearly four decades for $\delta = 0.38$. The difference in particle-size resolved relaxation times becomes even more extreme if one uses a more conventional definition of the structural relaxation time, e.g.\ the point where $\overline{F^{(s)}}(k,\tau_{\alpha}) = 1/e$. This  also implies that in the case of highly polydisperse mixtures, some caution is warranted when studying structural relaxation: on general grounds one expects the relaxation time to be defined as some point beyond which the correlator has decayed from its plateau by a certain amount, yet for the smallest particles the plateau height can become incredibly low. Hence, the meaning of `structural relaxation' may be qualitatively affected when using the definition ${F}_{\sigma}^{(s)}(k,\tau_{\alpha}) = 1/e$ (indicated by the horizontal black line in Fig.~\ref{fig:Fs_avg}), since for small particles this would probe only the short-time dynamics.

\begin{figure}[h]
    \centering
    \includegraphics[width=\columnwidth]{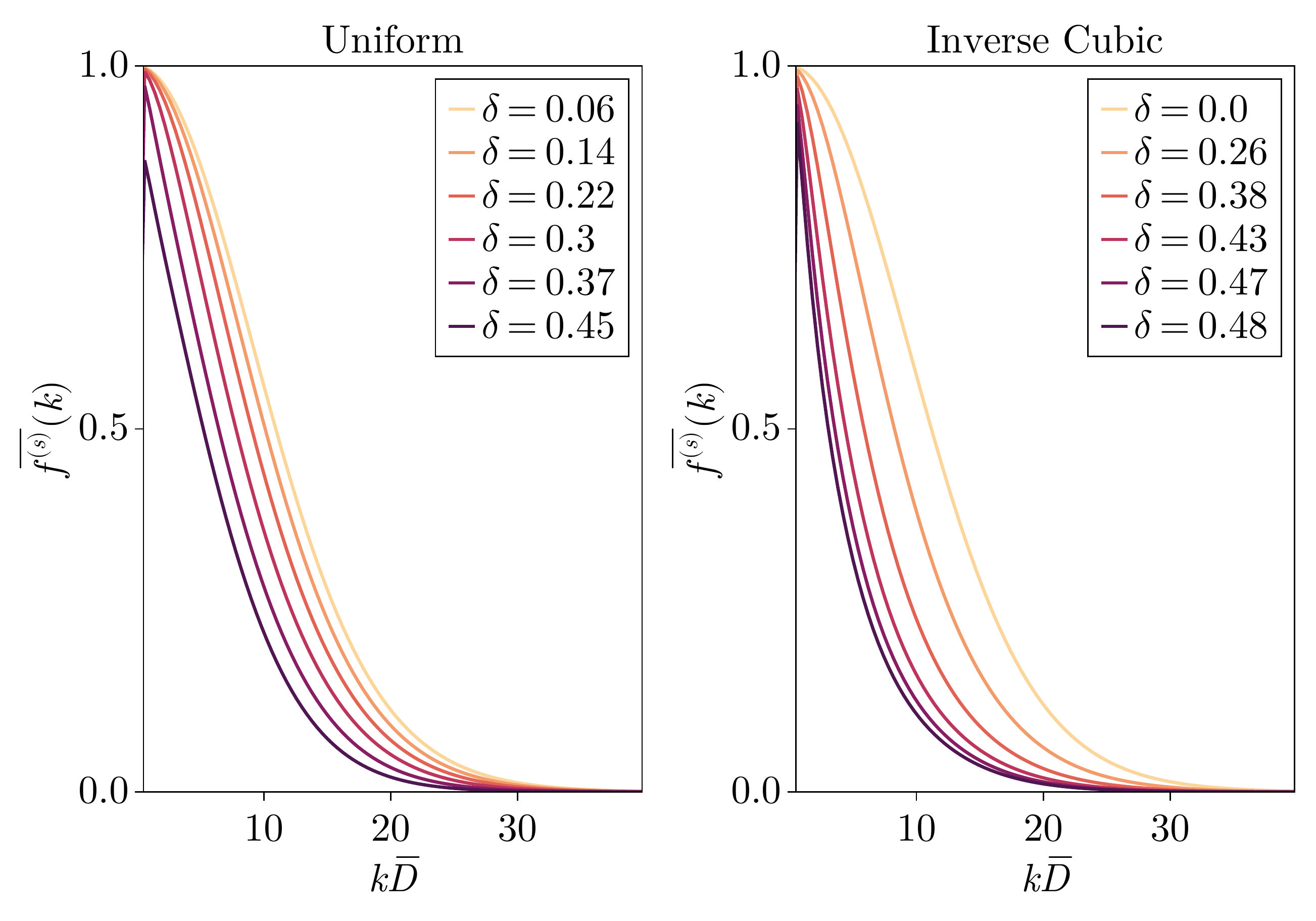}
    \caption{\textit{Diameter-averaged Lamb-M\"ossbauer factors determined from MCT for multi-component hard-sphere mixtures in the Percus-Yevick approximation as a function of dimensionless wave-number $k\overline{D}$. Left panel: uniform distribution with polydispersity index $\delta$; Right panel: inverse-cubic distribution with polydispersity index $\delta$}.}
    \label{fig:LM_PY}
\end{figure}

\begin{figure}[h]
    \centering
    \includegraphics[width=\columnwidth]{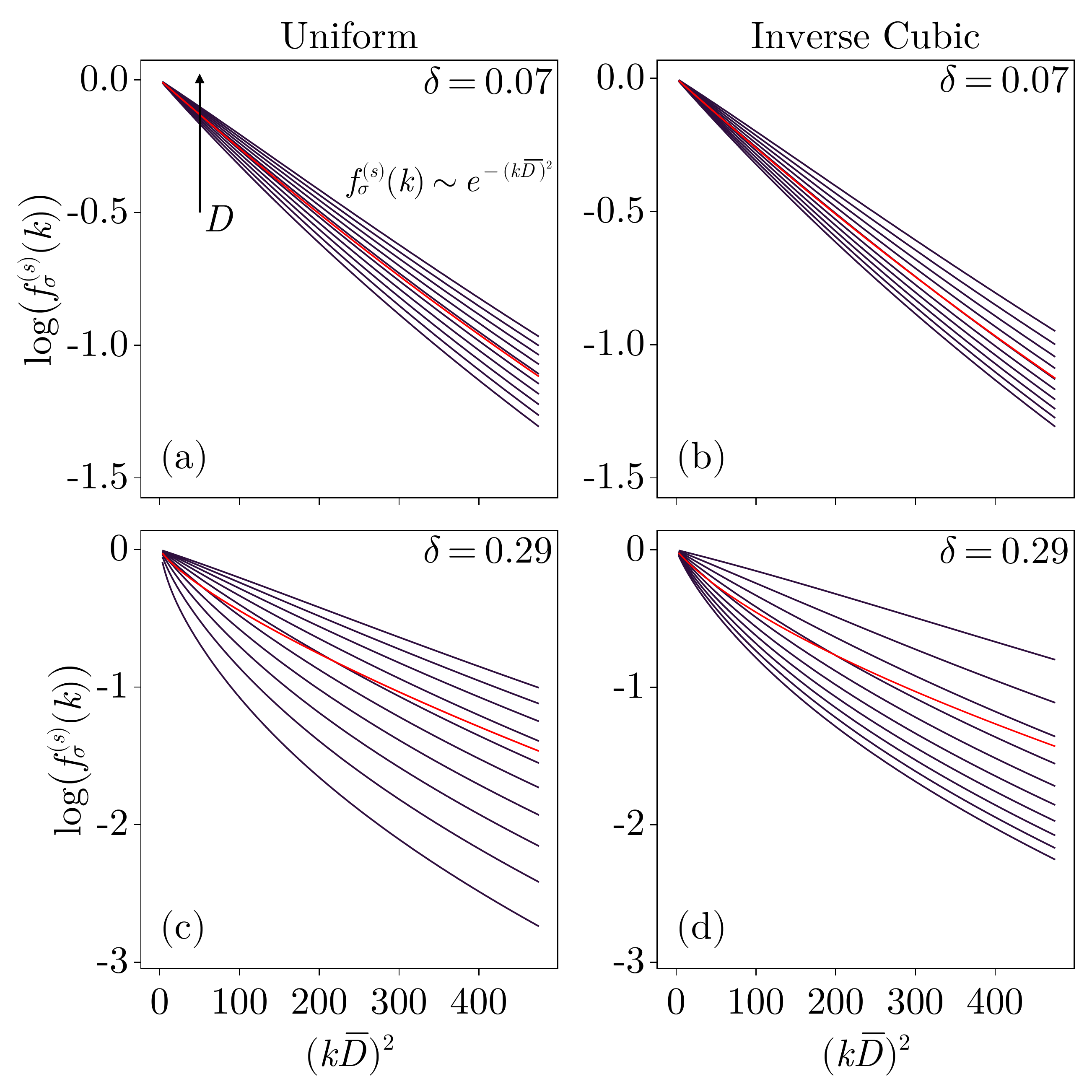}
    \caption{\textit{Diameter-resolved Lamb-M\"ossbauer factors determined from MCT for multi-component hard-sphere mixtures in the Percus-Yevick approximation as a function of the squared dimensionless wave-number $k\overline{D}$ on a log-linear scale. The red-line denotes the diameter-averaged Lamb-M\"ossbauer factor.}}
    \label{fig:log_LM}
\end{figure}

As can already be inferred from Fig.~\ref{fig:Fs_avg}, the Lamb-M\"{o}ssbauer factor, which is essentially the height of the plateau of the ISF, is highly species dependent. This indicates that local mobility is strongly dependent on particle diameter, since the inverse Fourier transform of the LM factor is related to the Van Hove function, i.e.\ the probability of observing a given displacement during an infinite observation window \cite{gotze2009complex}. More precisely, the half-width of the LM factor is inversely proportional to the root mean squared displacement of a given particle \cite{bengtzelius1984dynamics}, which we refer to as a localization length.  Let us first consider how the diameter-averaged form factors are affected by the polydispersity. In Fig.~\ref{fig:LM_PY} we show the species-averaged LM factors $\overline{f^{(s)}}(k) \equiv x_{\sigma}f_{\sigma}^{(s)}(k)$ evaluated at the MCT critical point for various values of the polydispersity index. For both distributions and for all wave-numbers, the LM factors decrease  monotonically with increasing $\delta$. Additionally, they decay over longer wave-lengths when the polydispersity index is increased, suggesting that the average localization length of all particles is increased by increasing size polydispersity. Furthermore, we find that the averaged LM factor is progressively less Gaussian with increasing $\delta$, which indicates that the local energy landscape around a given particle is progressively anharmonic \cite{gotze2009complex}.

To understand this, we argue that as the polydispersity index is increased, the notion of a local trapping cage becomes less relevant for the smallest particles. We show in Fig.~\ref{fig:log_LM} the diameter-resolved LM factors on a semi-logarithmic scale. For low degrees of polydispersity [panels (a)-(b)], we find the Gaussian result $f_{\sigma}^{(s)}(k) \propto \exp(-(k\overline{D})^2)$, which suggests that all particles effectively behave as if they were in a harmonic trap \cite{gotze2009complex}. As the polydispersity index is increased [panels (c)-(d)], we find that, while the largest particles retain a Gaussian LM factor, the smallest ones strongly deviate from one. Furthermore, the latter tend to decay faster as a function of $k$, signaling a much larger localization length. This supports the idea that the smallest particles are able to navigate heterogeneously (while still being localized) through narrow channels within the matrix of larger, more strongly localized ones. The red curves in Fig.~\ref{fig:log_LM} which represent the diameter-averaged LM factor demonstrate that at high polydispersity indices [panels (c), (d)], the non-Gaussian character of the diameter-averaged LM factor can be imputed to the smallest particles. These findings are consistent with earlier results on bidisperse mixtures \cite{thakur1991glass}, as well as experimental and computational work on continuously polydisperse systems \cite{zaccarelli2015polydispersity,heckendorf2017size}.

We find however that MCT is incapable of resolving another known observation in strongly polydisperse mixtures, namely that small particles escape their cage much earlier than large ones. Indeed, such reports have been made for simulated binary mixtures in the past \cite{flenner2005relaxation}, and more recently for the same inverse cubic distribution used in the first part of this work with $\delta \approx 0.23$ \cite{pihlajamaa2023polydispersity}. In these two studies, it was found that there is a dynamical separation between the smallest and the largest particles. Yet, MCT's asymptotic predictions are such that near the critical point considered here, species-resolved intermediate scattering functions decay from their respective plateaus at the same time, thus failing to capture the dynamical separation. This failure of MCT suggests that the theory effectively overestimates the coupling between different species, ultimately leading to an ideal glass transition scenario. This limitation is attributed to the general absence of activated events in the theory, which, if included, would manifest as additional relaxation channels, restoring ergodicity. The way in which these couple to the degree of polydispersity remains however an open question.

\subsection{Critical Dynamical Exponents}

Let us now focus on the critical exponents associated with the dynamical MCT scaling laws. We recall that these exponents are generally system-dependent, and therefore also depend on the chosen size distribution. However, given the qualitatively equivalent behavior of the uniform and inverse cubic size distributions above, we here consider only the case where the particle diameters follow a uniform distribution. Similar qualitative trends are expected to hold for other continuously polydisperse mixtures.

Although the critical scenario of MCT is unaffected by polydispersity (at least at reasonable degrees of polydispersity), we nonetheless observe important quantitative effects of polydispersity on the critical exponent $\gamma$ and the dynamical exponents $a, b$. We determine $\gamma$ by fitting the power law to the relaxation time $\overline{\tau}_{\alpha} \sim |\varphi-\varphi_c|^{-\gamma}$ for a set of numerical solutions in the range $10^{-2} \leq|\varphi-\varphi_c|\leq 10^{-5}$. The dynamical exponents $a,b$ are then determined by solving the system Eq.~\eqref{eq:MCT_exponent_relations} using standard root finding techniques.

\begin{figure}
\includegraphics[width=\columnwidth]{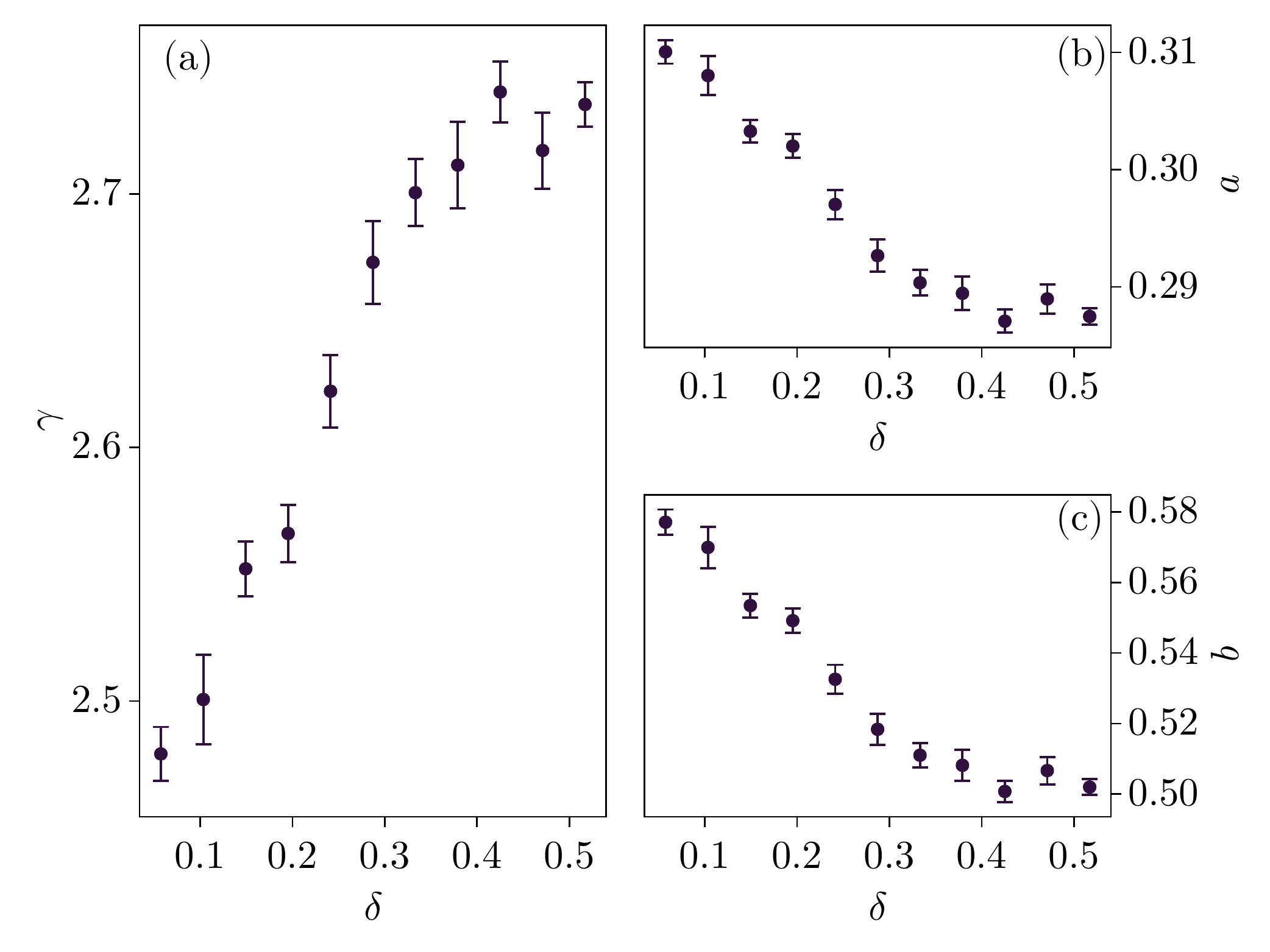}
\caption{(a) \textit{Critical exponent $\gamma$ as a function of the degree of polydispersity $\delta$. This exponent governs the divergence of the relaxation time.} (b) \textit{Dynamical exponent $a$, which governs the asymptotic behavior of the correlator when approaching the plateau from above.} (c) \textit{Dynamical exponent $b$ (the von Schweidler exponent) which governs the asymptotic behavior of the correlator when departing from the plateau. In the three panels, the error bars correspond to the uncertainty of the results of the fitting procedure (see Appendix \ref{app:fitting} for details)}.}
\label{fig:critical_exponents}
\end{figure}

We find that the exponent $\gamma$ increases monotonically with the level of polydispersity $\delta$, as shown in Fig.~\ref{fig:critical_exponents}(a). This increase implies a stronger divergence of the relaxation time, and thus, a higher degree of fragility with increasing polydispersity. Note here that we quantify the degree of fragility of a supercooled liquids as the deviation from an Arrhenius scaling of the relaxation time in terms of a power law, whereas fragility is generally measured from fits to the Vogel-Fulcher-Tammann equation \cite{debenedetti2001supercooled}. Concomitantly, the dynamical exponents $a$ and $b$ are found to decrease monotonically with $\delta$ [see Fig.~\ref{fig:critical_exponents}(b)-(c)]. The quantitative values of the critical exponents that we report here are also in line with earlier studies of polydisperse quasi-hard sphere mixtures \cite{weysser2010structural}. Moreover, we find that all exponents saturate around a polydispersity index of $\delta=0.4$, which coincides with the point at which the predicted value of $\varphi_c$ becomes constant for this size distribution (see Fig.~\ref{fig:phase_diagram}). These results further confirm that, within MCT, there is an upper bound for $\delta$ beyond which the dynamics are no longer affected by polydispersity for this distribution.

On general grounds, the critical exponents can be related to a dynamical length scale within the framework of inhomogeneous MCT \cite{biroli2006inhomogeneous}. Briefly, the growth of the (non-linear) dynamical susceptibility (i.e.\ the response of the coherent ISF to an infinitesimal localized perturbation of the density field near criticality) is governed by the values of the two dynamical exponents $a$ and $b$. The dynamical length scale, which measures the size of a strongly dynamically correlated region, grows as $\xi(t) \propto t^{a/2}$. Since $a$ decreases with $\delta$, this implies that increasing the polydispersity index hampers the growth of the dynamical length scale. This observation is in agreement with previous experimental findings \cite{abraham2008energy}, where it was found that lower polydispersities imply stronger dynamical heterogeneities. The above result is also consistent with the intuitive picture that we propose here: there exists a sub-population of particles with large localization lengths (the smallest ones), which fluidizes the system and thus makes it more difficult for the correlation length to grow in time, due to the inherent scrambling of the small particles in the voids surrounding  the largest ones.

\subsection{Stretched Exponential Relaxation}

It is well known that the coupling of different relaxation modes can lead to stretched exponential behavior \cite{phillips1996stretched}. In standard MCT, such stretching is usually attributed to the coupling of different wave-vectors, but the additional coupling of different particle species can lead to further stretching of the dynamics. Indeed, our species-averaged dynamics for the intermediate scattering functions (full lines in Fig.~\ref{fig:Fs_avg}), as well as the averaged von Schweidler exponent $b$ (Fig.~\ref{fig:critical_exponents}), indicate that additional relaxation channels provided by size polydispersity yield a stretching of the relaxation. 

\begin{figure}
\includegraphics[width=\columnwidth]{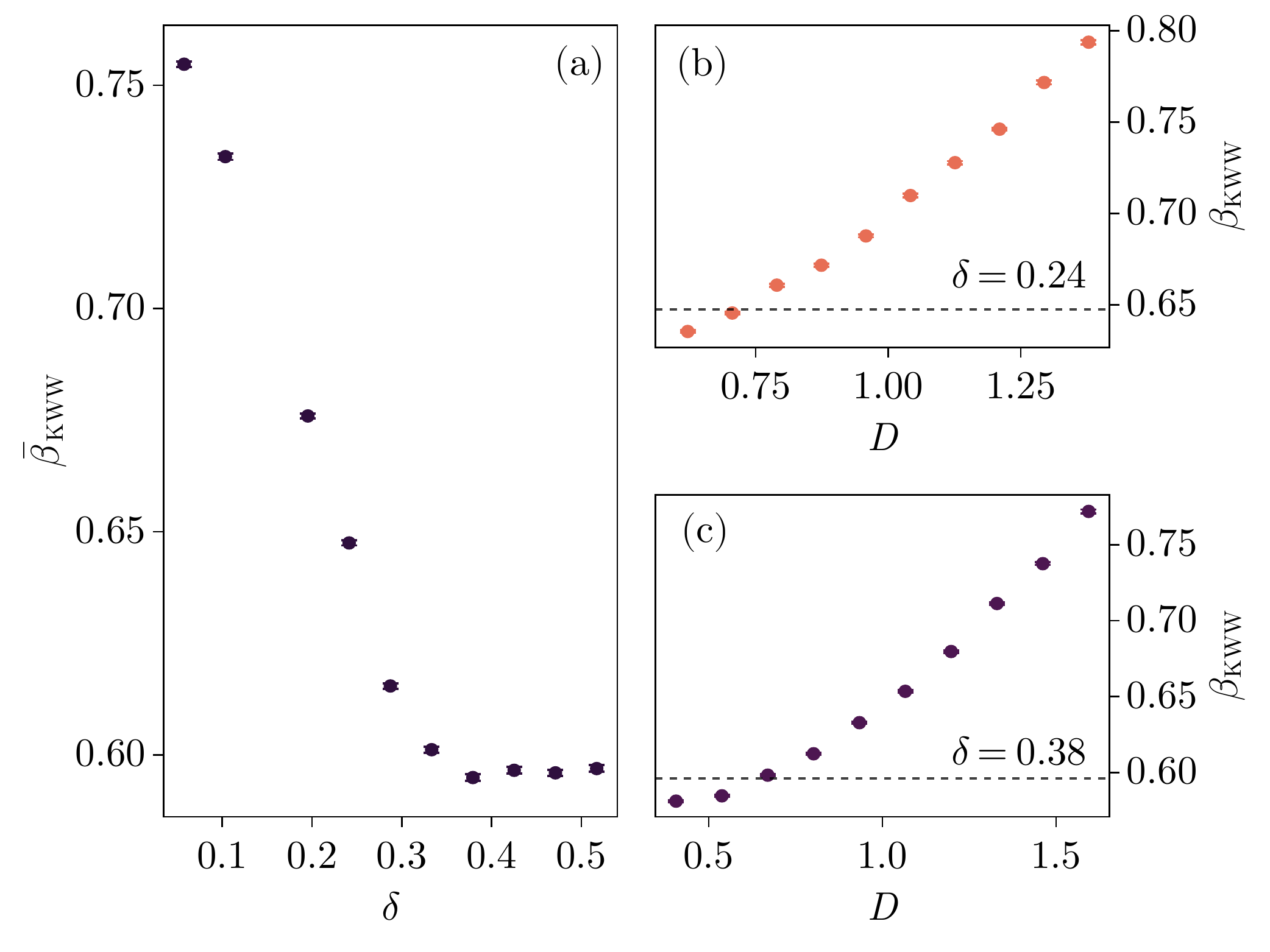}
\caption{(a) \textit{Kohlrausch–Williams–Watts exponent for the species-averaged incoherent intermediate scattering function $\overline{F}_s(k,t)$ as a function of the polydispersity index $\delta$.} (b)-(c) \textit{Kohlrausch–Williams–Watts exponent for species-specific incoherent intermediate scattering functions as a function of the particle diameter $D$, for polydispersity degree $\delta=0.24,\ 0.38$ respectively. The horizontal dashed line represents the  the species-averaged value $\overline{\beta}_{\text{KWW}}$.} \textit{ Error bars are smaller than the scatter point size.}}
\label{fig:KWW_analysis}
\end{figure}

In order to more quantitatively analyze the influence of polydispersity on the structural relaxation, we perform a detailed analysis of the long-time tail of the incoherent ISFs. Specifically, we fit a Kohlrausch function to the long-time decay of the species-averaged correlator $\overline{F^{(s)}}(k,t)$,
    \begin{equation}
        \overline{F^{(s)}}(k,t) = \overline{f^{(s)}}(k)\exp\left[-\left(\frac{t}{\overline{\tau}_{\alpha}}\right)^{\overline{\beta}_{\text{KWW}}}\right],
    \end{equation}
in order to extract the Kohlrausch–Williams–Watts (KWW) stretching exponent  $\overline{\beta}_{\text{KWW}}$ (see Appendix \ref{app:fitting} for details). Note that, within the context of MCT, this asymptotic form is strictly valid only in the infinite wave-number limit $k\rightarrow\infty$ \cite{fuchs1991comments, fuchs1994kohlrausch}, but it is also frequently employed to fit MCT predictions at finite wave-numbers \cite{kob1995testing, weysser2010structural} and it is a standard quantifier of relaxation in complex media \cite{phillips1996stretched}. Our results for $\overline{\beta}_{\text{KWW}}$ for a uniform size distribution are presented in Fig.~\ref{fig:KWW_analysis}(a). As the polydispersity degree is increased, we find that $\overline{\beta}_{\text{KWW}}$ decreases monotonically until it saturates beyond $\delta \approx 0.4$ at a low value around $\overline{\beta}_{\text{KWW}}\approx 0.60$. This saturation, which we recall is also observed in the state diagram and in the critical exponents, again strengthens the idea there is a degree of polydispersity beyond which MCT's prediction are no longer affected.

We subsequently perform a diameter-resolved analysis of the relaxation by computing the KWW exponent from the partial incoherent ISFs, which we denote $\beta_{\text{KWW}}$. The results are shown in Fig.~\ref{fig:KWW_analysis}(b) and (c) for $\delta = 0.24,\ 0.38$ respectively. The KWW exponent monotonically increases with the particle diameter in a continuous mixture. Furthermore, we find that the species-averaged exponent [dashed black lines in Fig.~\ref{fig:KWW_analysis} (b)-(c)] is very close to that of the value for the smallest particles. 

Since the stretching exponent is related to the degree of heterogeneity in relaxation timescales, it is not surprising that the diameter-averaged exponent is in fact governed by that of the more dynamically heterogeneous particles (i.e. the smallest ones). This interpretation is coherent with the notion that the growth rate of correlated regions is restrained by polydispersity, since the smallest particles are much more mobile than the largest ones. 

\subsection{Competing Relaxation Channels}

Lastly, we study the superposition of relaxation mechanisms induced by size polydispersity in a mixture with a uniform size distribution. To this end it is useful to translate our MCT results into frequency space. We follow standard procedure \cite{blochowicz2003susceptibility} and define the imaginary part of the response spectra $\operatorname{Im}[\overline{\chi}(k,\omega)] \equiv \overline{\chi}''(k,\omega)$ associated with $\overline{F^{(s)}}(k,t)$ as
\begin{equation}
    \overline{\chi}''(k,\omega) = -\int_{-\infty}^{\infty} d\log(t) \frac{d\overline{F^{(s)}}(k,t)}{d\log(t)} \frac{\omega t}{1+(\omega t)^2}, 
\label{eq:susceptibility_def}
\end{equation}
and analogously for the species-specific quantities, which we denote by $\chi''_{\sigma}(k,\omega)$. 

\begin{figure}[h!]
\includegraphics[width=\columnwidth]{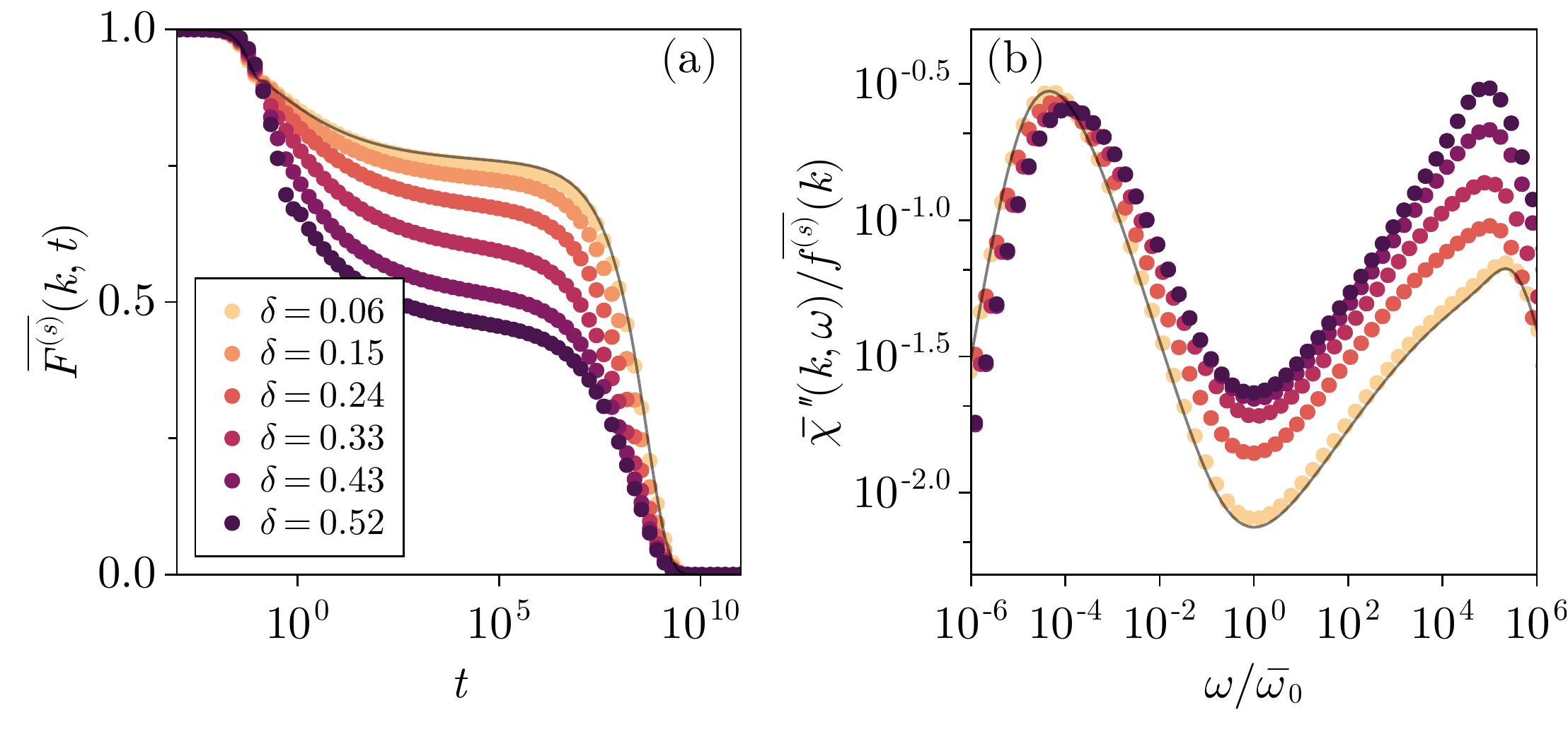}
\caption{(a) \textit{Species-averaged incoherent intermediate scattering functions obtained by numerically solving for Eq.~\eqref{eq:tagged_multicomponent_MCT} for polydispersity degree $\delta$.} (b) \textit{Corresponding susceptibility spectra $\overline{\chi}_s''(k,\omega)$ rescaled by the averaged Lamb-M\"{o}ssbauer factor. Note that the x-axis is rescaled by the frequency $\overline{\omega}_0$ at which the minimum of the spectrum occurs. The solid line corresponds to the single-component MCT result in both panels.}}
\label{fig:susceptibility_spectra}
\end{figure}

Figure~\ref{fig:susceptibility_spectra}(b) shows the response spectra for different polydispersity degrees at a fixed relaxation time $\overline{\tau}_{\alpha}$, rescaled by the average LM factor such that the low-frequency peaks (associated with structural $\alpha$-relaxation) all collapse at the same height. Note that minor deviations from a perfect collapse are noticeable, which are attributed to our numerical accuracy in determining solutions at fixed relaxation time.   
It can be seen that the $\alpha$-peak broadens toward higher frequencies as the polydispersity index $\delta$ increases. This fanning is associated with an increased stretching of the long-time structural relaxation of $\overline{F^{(s)}}(k,t)$, in line with the results of Fig.~\ref{fig:KWW_analysis}(a) and explicitly shown in Fig.~\ref{fig:susceptibility_spectra}(a). 
Compared to the monodisperse case (full line in Fig.~\ref{fig:susceptibility_spectra}), we infer that polydispersity induces additional relaxation channels that now compete with each other, thus resulting in a significantly broader spectrum [see Fig.~\ref{fig:susceptibility_spectra}(b)]. 
We also find that as the degree of polydispersity increases, the relative height of the boson peak (i.e.\ the high-frequency peak of the spectrum) increases with respect to that of the $\alpha$-peak. Since the two spectral peaks are separated by the caging regime (i.e.\ the minimum of the spectrum), this further corroborates that the degree of polydispersity influences the ratio between structural relaxation before and after the caging regime. In particular, for the highest polydispersity index considered in this work, the magnitude of the boson peak exceeds that of the principal $\alpha$-relaxation, which implies in the time domain [Fig.~\ref{fig:susceptibility_spectra}(a)] that more than 50\% of the decay in density correlations occurs before reaching the caging plateau. 

\begin{figure}
\includegraphics[width=\columnwidth]{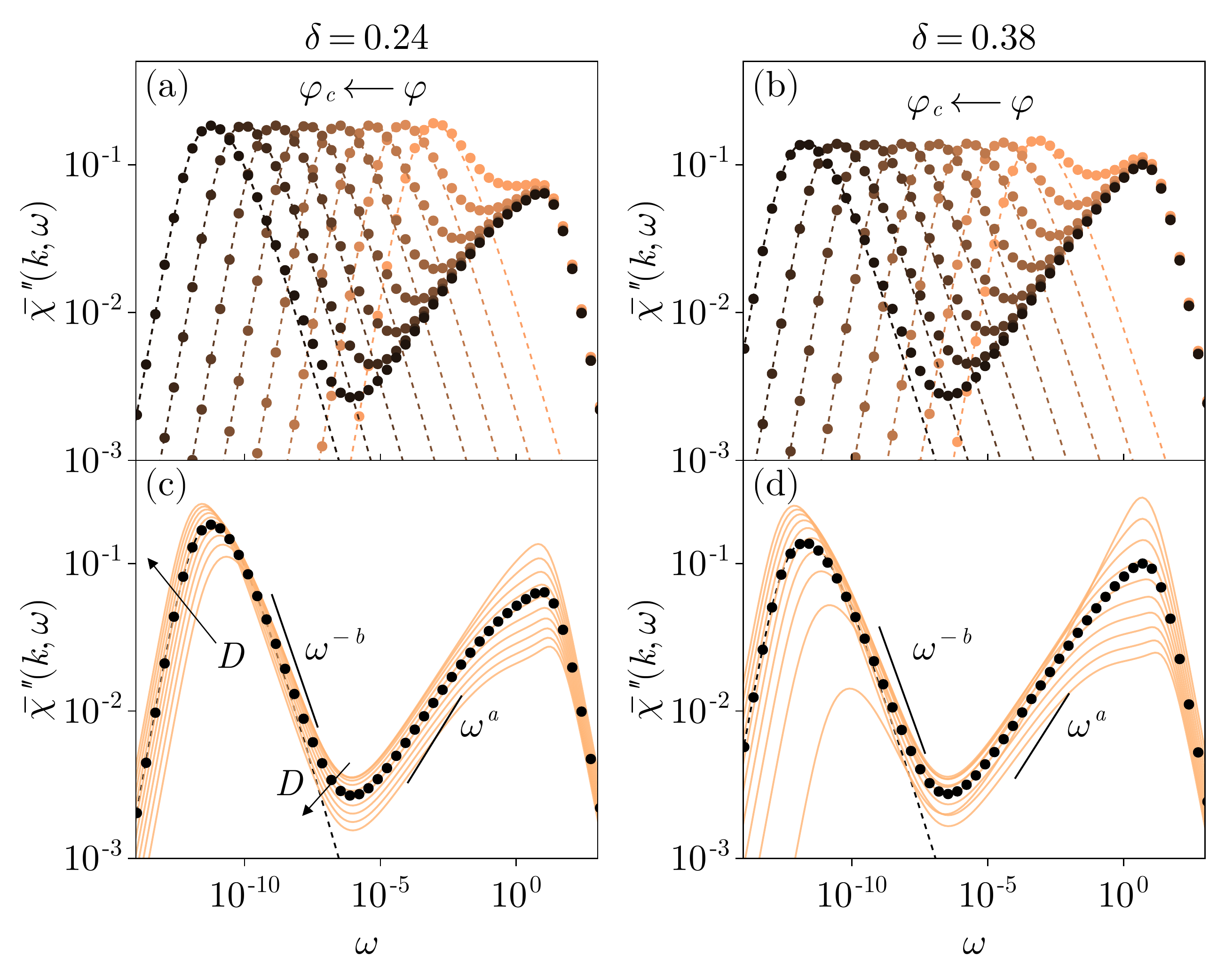}
\caption{Panels (a), (b): \textit{Species-averaged susceptibility spectra on a double logarithmic scale as the critical packing fraction $\varphi_c$ is approached from below for polydispersity degree $\delta = 0.24,\ 0.38$ respectively. Dashed lines correspond to the spectrum of a Kohlrausch fit for the diameter averaged susceptibility $\overline{\chi}''(k,\omega)$, see Appendix \ref{app:fitting} for details.} Panels (c), (d): \textit{Species-averaged susceptibility (scatter points) and species-resolved (colored) on a double logarithmic scale at a distance $|\varphi-\varphi_c| = 10^{-5}$ to the critical point. Dashed black lines correspond to the spectrum of a Kohlrausch fit for the diameter averaged susceptibility $\overline{\chi}''(k,\omega)$, see Appendix \ref{app:fitting} for details. The arrows indicate the direction of increasingly large particle diameters $D$.}}
\label{fig:susceptibility_spectra_epsilon}
\end{figure}

Let us now inspect the $\alpha$-peak of the spectrum more closely. The dashed lines in Fig.~\ref{fig:susceptibility_spectra_epsilon}(a), (b) represent the spectrum of the fitted stretched exponentials at polydispersities $\delta = 0.24,\ 0.38$ for varying packing fractions towards the critical point. For all cases, we find that the signal is well captured up to the vicinity of the spectral minimum, where a smooth transition occurs between  stretched exponential relaxation and the \textit{von Schweidler excess}: $\overline{\chi}''(\omega) \sim \omega^{-b}$ [shown for clarity in panels (c) and (d)]. Note however that this transition is hard to see, since the von Schweidler exponent and the slope of the stretched exponential are very similar. In order to understand the microscopic origin of the excess in the relaxation spectrum, we study the diameter-resolved susceptibility spectra, shown in Fig.~\ref{fig:susceptibility_spectra_epsilon}(c) and (d) by the colored lines. Here it is important to recall that the asymptotic properties of MCT around the minimum of the spectra are universal. In particular, the theory predicts that the von Schweidler exponent $b$ for the species-resolved susceptibilities $\chi_{\sigma}''(k,\omega \rightarrow\omega_0^{-}) \sim K^{(-)}_{\sigma}(k)\omega^{-b}$ is diameter independent \cite{franosch1997asymptotic, voigtmann2003mode} [although it is system dependent, as previously shown in Fig.~\ref{fig:critical_exponents}(c)]. The results of Fig.~\ref{fig:susceptibility_spectra_epsilon}(c) and (d) reveal that the particle-resolved susceptibilities (full colored lines) indeed all share the same power laws, but the location of the low-frequency $\alpha$-peak is however diameter dependent. To the left of the $\alpha$-peak, the highest curve corresponds to that of the largest diameter while the lowest curve to the smallest diameter. At the minimum $\overline{\omega}_0$ the reverse situation is found, as indicated by the black arrows pointing towards increasing diameters. Hence, the curves must cross at some point beyond which the smallest particles contribute `in excess' to the spectrum.

In fact, MCT guarantees that the distance between the  $\alpha$-peaks of particles with different diameters is fixed as one approaches the transition, since they all share the same critical exponent $\gamma$. Hence, as we get closer to the critical point the spectral contributions from the smallest particles at frequencies past the caging regime do not increase in magnitude, and the entire effect remains subtle. However in molecular dynamics simulations, we know that this is not valid, and that in fact, beyond the spurious transition predicted by MCT, the distance between the $\alpha$-peaks of particles with different diameters must increase, since the time interval between which the smallest and the largest particles relax increases with deeper supercooled systems \cite{pihlajamaa2023polydispersity}. This raises an important question regarding the exact nature and the true microscopic origins of relaxational excess observed in the spectrum of simple, yet polydisperse glass formers. In particular it would be of interest to compare the value of the excess wing exponent reported in \cite{guiselin2022microscopic} with that of the von-Schweidler exponent for the same system, and to perform a similar diameter-resolved analysis of the relaxation spectra in order to check whether or not there is a homogeneous spectral contribution with respect to particle diameters. 

\section{Conclusion}
In this work, we have analyzed the dynamics of continuously polydisperse glass-forming liquids through the lens of Mode-Coupling Theory. We have demonstrated that increasing the degree of size polydispersity first slows down the dynamics, before pushing the critical point $\varphi_c$ to higher packing fractions, thus stabilizing the liquid phase. Of the three size distributions that we have considered, the liquid stabilization effect is largest for the inverse cubic distribution. This generalizes earlier MCT results on simpler, binary mixtures. A detailed study of the Lamb-M\"{o}ssbauer factors has revealed that the smaller particles have much longer localization lengths than the larger ones, and strongly deviate from the standard Gaussian behavior. This leads to visible deviations from Gaussian behavior also in diameter-averaged quantities.

Our analysis of the particle-averaged dynamics near the critical point shows that increasing the degree of polydispersity induces a significant stretching of the KWW exponent, a lowering of the critical dynamical exponents $a,\ b$, and a net growth of the critical exponent $\gamma$ that governs the divergence of the relaxation time. Our results thus indicate that increasing the polydispersity index affects the fragility of the liquid, as well as dynamical heterogeneities whose growth is dictated by the dynamical exponents $a$ and $b$ \cite{biroli2006inhomogeneous, szamel2010diverging}.

By performing a diameter-resolved analysis, we have shown that the Kohlrausch–Williams–Watts stretching exponent of the diameter-averaged dynamics is dominated by the dynamics of the smaller particles, which have significantly lower Kohlrausch–Williams–Watts exponents compared to their larger counterparts. This leads us to infer that temporal heterogeneities in the dynamics are also mainly due to the smaller particles. This conclusion also is in partial agreement with recent observations in molecular dynamics simulations \cite{pihlajamaa2023polydispersity}, where it has been shown that smaller particles have strongly heterogeneous dynamics compared to their larger counterparts. We have further studied the manifestation of this effect in the susceptibility spectra of the relaxation, which exhibit a broadening and von Schweidler excess that becomes more pronounced upon increased polydispersity. We can attribute this broadening to two factors: firstly, to the increasing temporal heterogeneities of the small particles as they exhibit low stretching exponents (compared to the larger ones), and secondly to the shift in spectral peak positions of the smallest particles. It is important to note, however, that the von Schweidler excess remains strictly governed by the critical exponent $b$, but since $b$ itself changes with polydispersity, so does the spectrum. A detailed diameter-resolved investigation using computer simulations could verify if these observations persist beyond the MCT regime.

Overall, our results demonstrate that size polydispersity imposes important and non-trivial effects on glassy dynamics. This is particularly relevant in the context of deeply supercooled liquids that typically require a large degree of polydispersity in order to permit equilibration at temperatures below the MCT crossover temperature. Our work reveals, on purely theoretical grounds, that even in the mildly supercooled regime where MCT is usually deemed applicable, the presence of polydispersity can induce complex dynamical features. It is possible that some of these polydispersity-specific features might carry over into the more deeply supercooled regime. 

Furthermore, we have shown that strongly continuously polydisperse mixtures of hard spheres display features of both collective glassy behavior for the largest particles, and localization in narrow channels for the smallest ones. In this sense, such systems provide an interesting intermediate step between effectively monodisperse glasses and the more minimal model of the random Lorentz gas used to study transport in heterogeneous environments \cite{hofling2006localization}, which is governed by percolation in physically relevant dimensions.

\begin{acknowledgments}
We thank Thomas Voigtmann \& Vincent Debets for insightful discussions. Vincent Debets \& Chengjie Luo are also gratefully acknowledged for their critical reading of the manuscript. This work has been financially supported by the Dutch Research Council (NWO) through a Vidi grant (IP, CCLL, and LMCJ) and START-UP grant (LMCJ). \\
\end{acknowledgments}

\textbf{Author Contributions}: CCLL, IP, and LMCJ designed the research. CCLL performed the research and wrote the manuscript. IP developed a majority of the numerical routines. IP and LMCJ revised the manuscript.

\appendix

\section{Validity of Effective Discrete Representations for Continuously Polydisperse Hard-Spheres}
\label{app:quantisation}

\begin{figure}[h]
    \centering
    \includegraphics[width=\columnwidth]{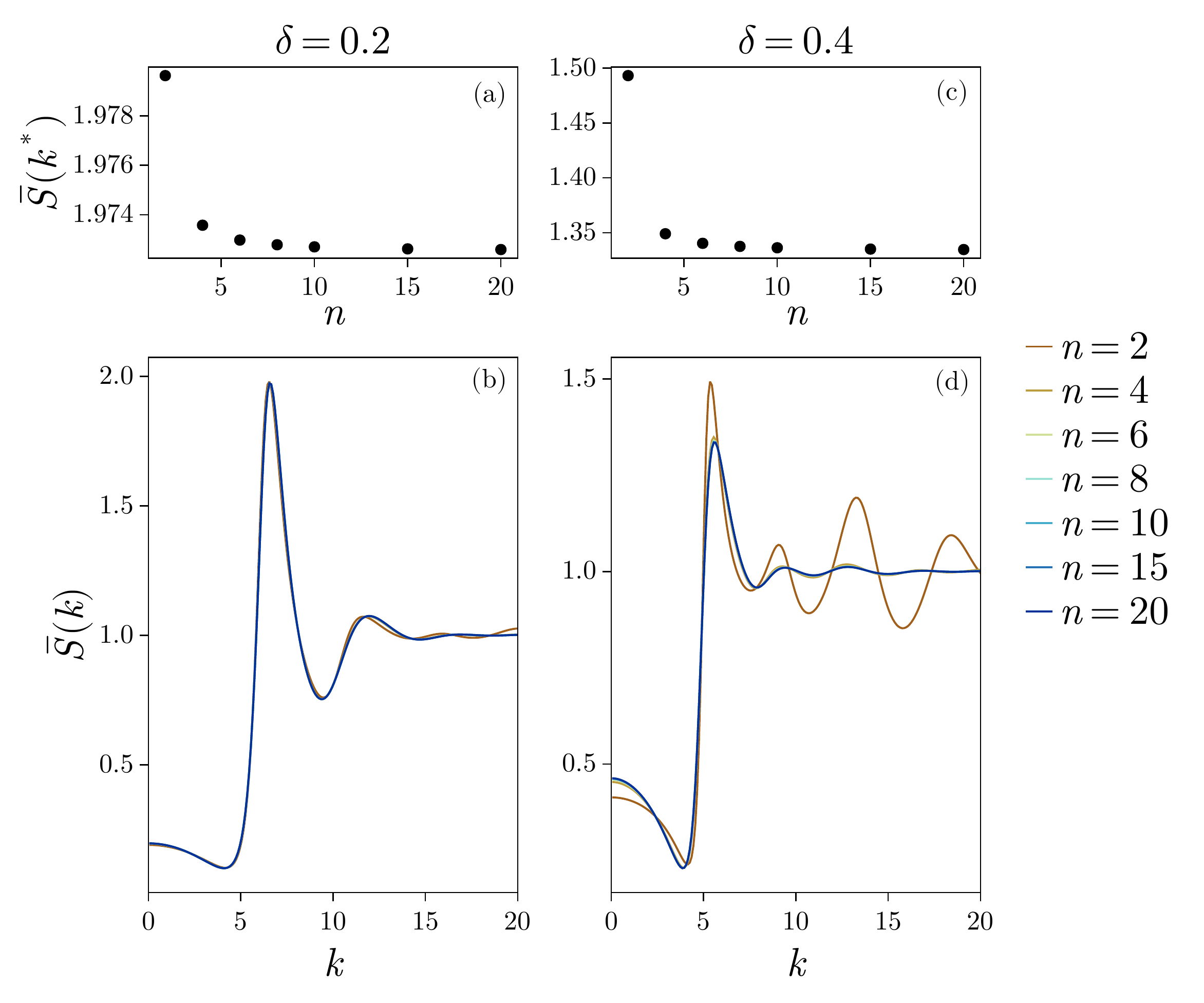}
    \caption{(a)-(b) \textit{Convergence of the static structure factor for a uniform distribution of diameters at fixed polydispersity $\delta = 0.2$ and packing fraction ($\varphi = 0.521$) of a discrete description of the continuously polydisperse system.} (c)-(d) \textit{Same as panels} (a)-(b) \textit{for a polydispersity index of $\delta=0.4$}.}
    \label{fig:Sk_convergence}
\end{figure}

We first discuss the validity of the discrete representation of a continuous distribution of particle sizes. We show in Fig.~\ref{fig:Sk_convergence} the structure factors for a uniform continuous distribution of particle diameters, as determined from multi-component Percus-Yevick hard-spheres, for two degrees of polydispersity $\delta= 0.2,\ 0.4$ as a function of an effective $n$-component representation at fixed packing fraction $\varphi=0.521$. The top panels show the convergence of the peak of the averaged structure factor $\overline{S}(k^*)$ as a function of the number of resolved components. We judge that an effective 10-component description is sufficient to accurately resolve the total structure of the system, motivating the choice in the main text. Similar results are found for other degrees of polydispersity, packing fractions, as well as for other type of distributions.

\begin{figure}
    \centering
\includegraphics[width=\columnwidth]{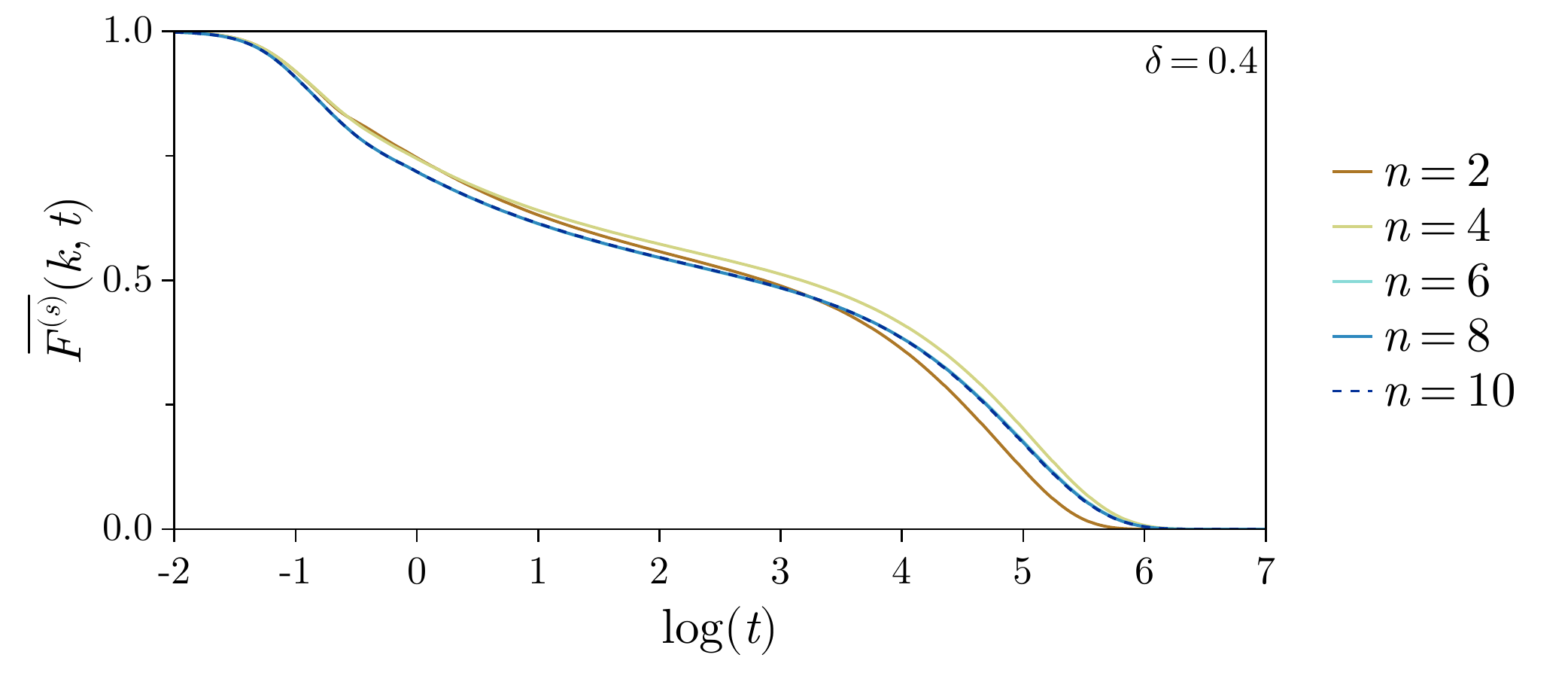}
    \caption{\textit{Convergence of the diameter averaged incoherent intermediate scattering function $\overline{F^{(s)}}(k,t)$ as a function of the number of resolved diameters $n$ for a polydispersity index $\delta=0.4$ at packing fraction $\varphi=0.521$.}}
    \label{fig:Fk_convergence}
\end{figure}

\begin{figure}
    \centering
\includegraphics[width=\columnwidth]{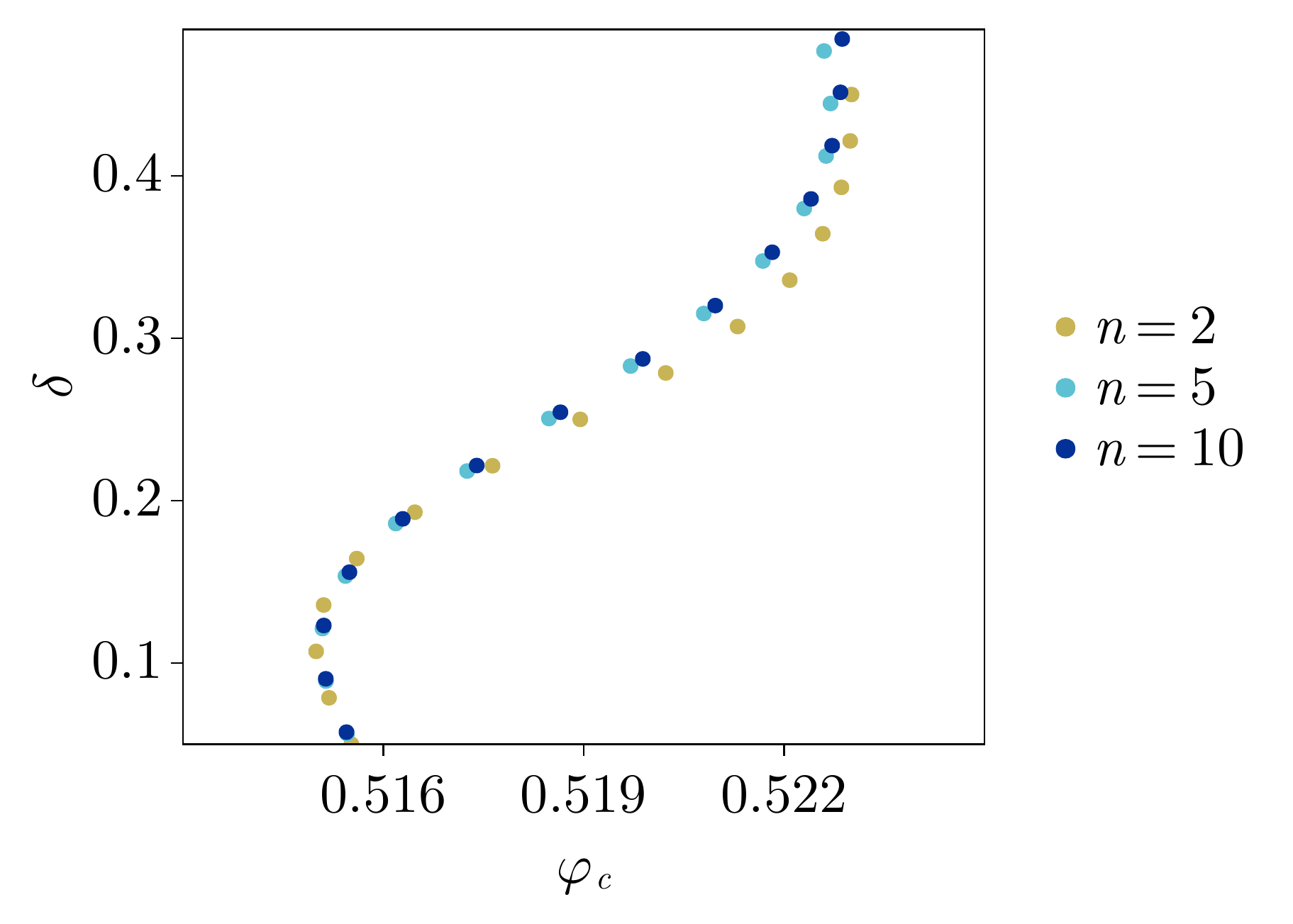}
    \caption{\textit{Convergence of the phase diagram for a uniform distribution of diameters as a function of the number of resolved components $n$.}}
    \label{fig:phase_diagram_convergence}
\end{figure}

To further illustrate this point we plot in Fig.~\ref{fig:Fk_convergence} the convergence of the resulting MCT solutions as a function of the number of resolved species for the diameter-averaged incoherent intermediate scattering function. We find that the curves for $n=6,\ 8\text{ and } 10$ effective components have converged to the same value. To make the convergence manifest we have dashed the $n=10$ curve. Similar results are found for other degrees of polydispersity and other ranges of packing fractions.
Lastly we show in Fig.~\ref{fig:phase_diagram_convergence} that the qualitative picture of the phase diagram is independent of the number of resolved diameters for a uniform distribution of diameters. Similar results are found for the other distributions.

\section{Numerical Details}
\label{app:numerical_details}

\begin{figure}[h]
    \centering
    \includegraphics[width=0.8\columnwidth]{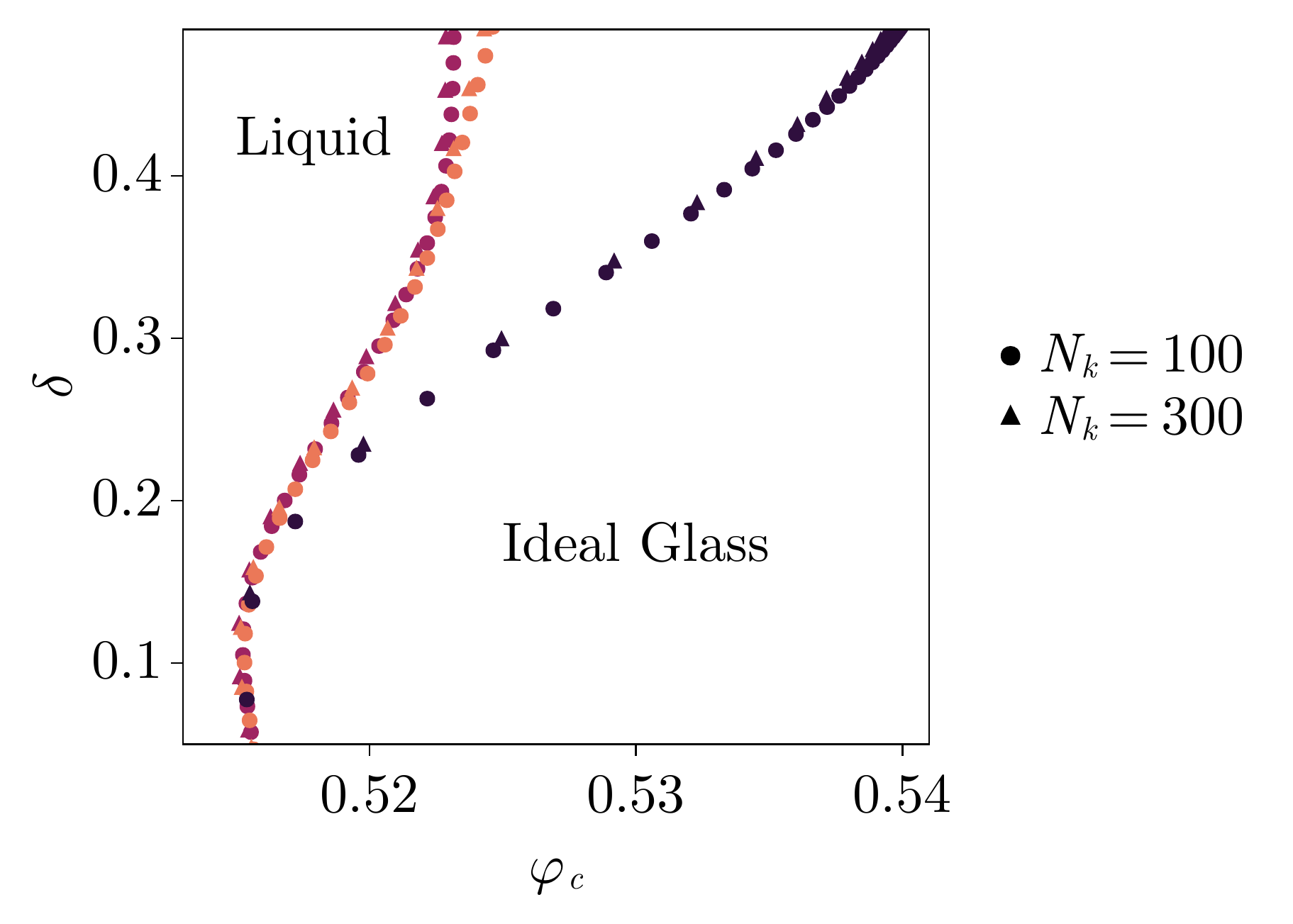}
    \caption{\textit{Phase diagram computed with two different wave-vector grids. Circle scatter points correspond to those shown in the main text. The upwards facing triangles correspond to a wave-vector grid with twice the resolution, and resolved over smaller wave-lengths.}}
    \label{fig:phase_diagram_fine_grid}
\end{figure}

The equations of motion \eqref{eq:tagged_multicomponent_MCT}-\eqref{eq:collective_multicomponent_MCT} were solved using a time-doubling algorithm \cite{fuchs1991comments,flenner2005relaxation}, which has been recently been implemented in an open-source solver for integro-differential equations of the mode-coupling type \cite{MCT_solver}. The wave-vector integrals were performed (in bi-polar coordinates) over a grid of $N_k=100$ equidistant points between $k_{\text{min}} = 0.2$ and $k_{\text{max}} = 39.8$. We have checked the stability of our results against a finer grid with $k_{\text{min}} = 0.2$ and $k_{\text{max}} = 60.0$, with $N_k=300$ equidistant points in between (which doubles the resolution compared to the one presented in the main text). Results are shown in Fig.~\ref{fig:phase_diagram_fine_grid}; we observe no qualitative differences between the two wave-vector grids. 

\section{Static Structure Factors}
\label{app:structure}
\begin{figure}
    \centering
    \includegraphics[width=\columnwidth]{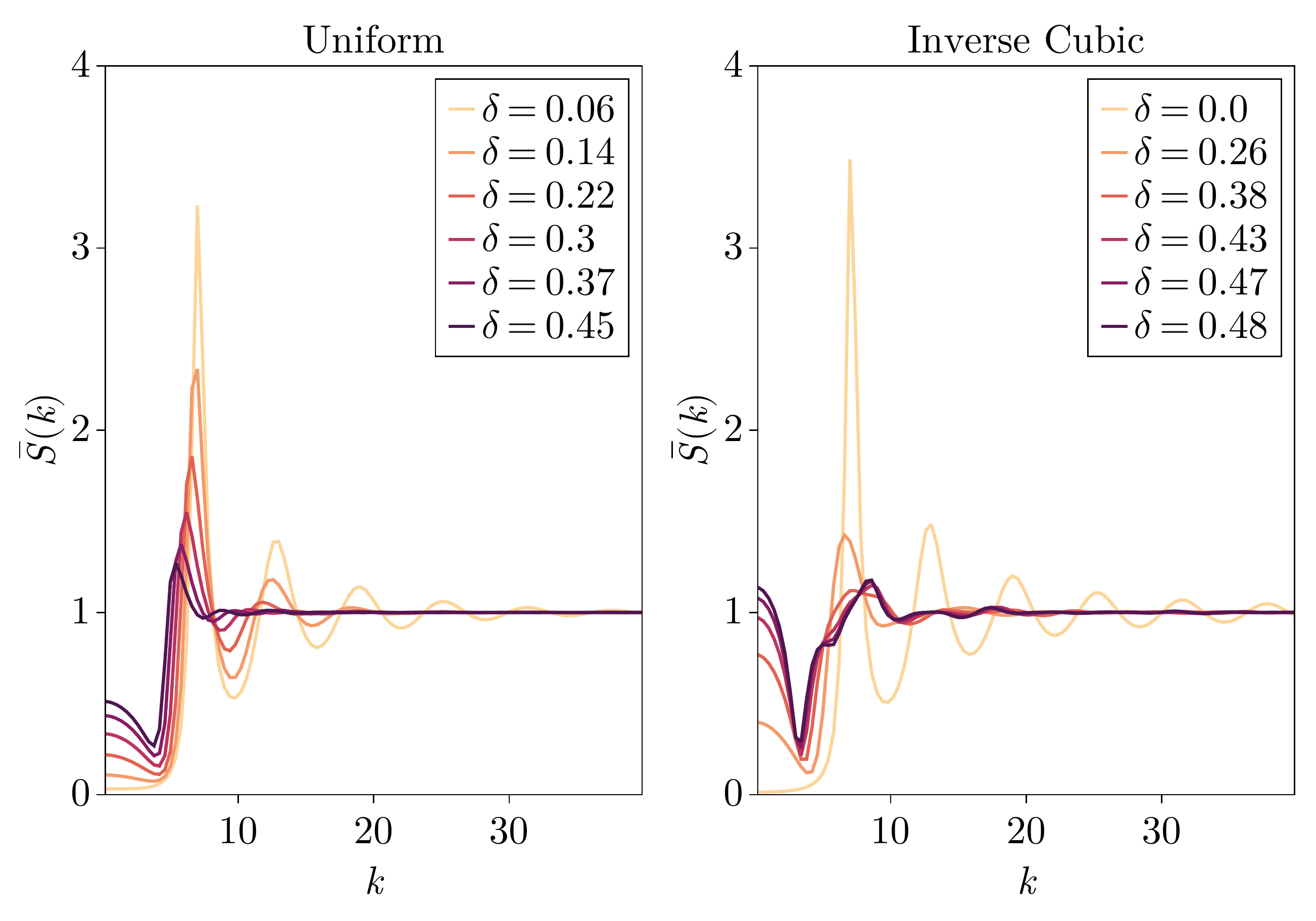}
    \caption{\textit{Diameter-averaged static structure factors for effective 10-component mixtures with polydispersity $\delta$ as determined from Percus-Yevick hard-spheres. Left: uniform distribution of particle diameters, right: inverse cubic distribution.}}
    \label{fig:struct_fact}
\end{figure}

We show in Fig.~\ref{fig:struct_fact} the averaged static structure factor $\overline{S}(k)$, as determined from multi-component Percus-Yevick hard-spheres, for various polydispersity indices, evaluated at the critical point $\varphi_c$ for an effective 10-component continuously polydisperse mixture. For both distributions, as the degree of polydispersity is increased, we see that the magnitude of the first peak of the averaged structure factor drastically diminishes. In the case of the uniform distribution, we find that the peak is modestly shifted to lower wave-numbers, signaling a slight increase in the average cage size. We see however that in the case of the inverse cubic distribution, the structure is completely washed out as the polydispersity index is increased, and in fact the qualitative aspect of the structure factor is significantly changed by the abundance of very small particles.

\section{Fitting Procedures}
\label{app:fitting}
\subsubsection*{Kohlsrauch Law}
We fit the stretched exponential Kohlrausch law for the averaged, and species resolved incoherent intermediate scattering functions when they lie between half of and one-tenth of the corresponding Lamb-M\"{o}ssbauer factor, i.e. $C(t) \in [C_{\infty}/2, C_{\infty}/10]$ for a correlation function $C(t)$ exhibiting a plateau at $C_{\infty}$. This ensures that we are well away from the MCT von Schweidler decay, but also in a range that is comparable with what would generally be done for experimental and simulation results, as the long time tails have a tendency to be statistically unreliable. We note that in the case where the plateau is extremely low, as is the case for the smallest particles in the very high polydispersity limit (see Fig.~\ref{fig:Fs_avg}-(c)), we cannot reasonably perform the fit. \\

We would like to add that fitting a stretched exponential decays to correlation functions is a complicated matter. This is partly due to the fact that on general grounds, such an asymptotic form cannot be rigorously justified. However, this does not mean that the KWW exponent extracted is not useful: it is a quantifier of relaxation that allows for an easy comparison in many different systems. We simply wish to emphasize that the quantitative values of the KWW exponent are very sensitive to the fitting interval. To address this, we have checked that various fitting intervals (i.e. including a reasonable larger portion of the tail) does not affect the qualitative picture we presented, although it does affect the quantitative picture.

\subsubsection*{Critical Exponents}
In order to extract the critical exponents $a, b$ from the MCT solutions, we use a set of solutions at distance  $ |\varphi-\varphi_c| \in [10.0^{-6}, 10.0^{-5}]$ for the distance to the critical packing fraction $\varphi_c$, and determine the exponent of the diverging relaxation time $\gamma$, which satisfies $\tau_{\alpha} \propto |\varphi-\varphi_c|^{-\gamma}$. We then make use of the two following identities which can be rigorously derived from the theory \cite{bengtzelius1984dynamics,gotze2009complex}: $\gamma = 1/2a + 1/2b$ and $\Gamma(1-a)^2/\Gamma(1-2a) = \Gamma(1+b)^2/\Gamma(1+2b)$ which can be numerically solved for the exponents $a$ and $b$ using standard root finding methods.

\subsubsection*{Spectral Susceptibility}
We compute the spectral susceptibility $\chi''(k,\omega)$ (i.e. the imaginary part of the dynamical compressibility) by integrating Eq.~\eqref{eq:susceptibility_def} numerically over a logarithmic scale using a trapezoidal rule over a range of times exceeding complete structural relaxation, to ensure appropriate convergence. We have checked that more advanced integration methods yield the same results (i.e. Gauss–Kronrod quadrature). The time derivative of the intermediate scattering function in the integrand was obtained using spline interpolation.

\bibliography{apssamp}

\providecommand{\noopsort}[1]{}\providecommand{\singleletter}[1]{#1}%
\begin{thebibliography}{58}%
\makeatletter
\providecommand \@ifxundefined [1]{%
 \@ifx{#1\undefined}
}%
\providecommand \@ifnum [1]{%
 \ifnum #1\expandafter \@firstoftwo
 \else \expandafter \@secondoftwo
 \fi
}%
\providecommand \@ifx [1]{%
 \ifx #1\expandafter \@firstoftwo
 \else \expandafter \@secondoftwo
 \fi
}%
\providecommand \natexlab [1]{#1}%
\providecommand \enquote  [1]{``#1''}%
\providecommand \bibnamefont  [1]{#1}%
\providecommand \bibfnamefont [1]{#1}%
\providecommand \citenamefont [1]{#1}%
\providecommand \href@noop [0]{\@secondoftwo}%
\providecommand \href [0]{\begingroup \@sanitize@url \@href}%
\providecommand \@href[1]{\@@startlink{#1}\@@href}%
\providecommand \@@href[1]{\endgroup#1\@@endlink}%
\providecommand \@sanitize@url [0]{\catcode `\\12\catcode `\$12\catcode
  `\&12\catcode `\#12\catcode `\^12\catcode `\_12\catcode `\%12\relax}%
\providecommand \@@startlink[1]{}%
\providecommand \@@endlink[0]{}%
\providecommand \url  [0]{\begingroup\@sanitize@url \@url }%
\providecommand \@url [1]{\endgroup\@href {#1}{\urlprefix }}%
\providecommand \urlprefix  [0]{URL }%
\providecommand \Eprint [0]{\href }%
\providecommand \doibase [0]{https://doi.org/}%
\providecommand \selectlanguage [0]{\@gobble}%
\providecommand \bibinfo  [0]{\@secondoftwo}%
\providecommand \bibfield  [0]{\@secondoftwo}%
\providecommand \translation [1]{[#1]}%
\providecommand \BibitemOpen [0]{}%
\providecommand \bibitemStop [0]{}%
\providecommand \bibitemNoStop [0]{.\EOS\space}%
\providecommand \EOS [0]{\spacefactor3000\relax}%
\providecommand \BibitemShut  [1]{\csname bibitem#1\endcsname}%
\let\auto@bib@innerbib\@empty
\bibitem [{\citenamefont {Kawasaki}\ and\ \citenamefont
  {Tanaka}(2011)}]{kawasaki2011structural}%
  \BibitemOpen
  \bibfield  {author} {\bibinfo {author} {\bibfnamefont {T.}~\bibnamefont
  {Kawasaki}}\ and\ \bibinfo {author} {\bibfnamefont {H.}~\bibnamefont
  {Tanaka}},\ }\bibfield  {title} {\bibinfo {title} {Structural signature of
  slow dynamics and dynamic heterogeneity in two-dimensional colloidal liquids:
  Glassy structural order},\ }\href@noop {} {\bibfield  {journal} {\bibinfo
  {journal} {J. Phys. Condens. Matter.}\ }\textbf {\bibinfo {volume} {23}},\
  \bibinfo {pages} {194121} (\bibinfo {year} {2011})}\BibitemShut {NoStop}%
\bibitem [{\citenamefont {Berthier}\ and\ \citenamefont
  {Reichman}(2023)}]{berthier2023modern}%
  \BibitemOpen
  \bibfield  {author} {\bibinfo {author} {\bibfnamefont {L.}~\bibnamefont
  {Berthier}}\ and\ \bibinfo {author} {\bibfnamefont {D.~R.}\ \bibnamefont
  {Reichman}},\ }\bibfield  {title} {\bibinfo {title} {Modern computational
  studies of the glass transition},\ }\href@noop {} {\bibfield  {journal}
  {\bibinfo  {journal} {Nat. Rev. Phys.}\ ,\ \bibinfo {pages} {1}} (\bibinfo
  {year} {2023})}\BibitemShut {NoStop}%
\bibitem [{\citenamefont {Ninarello}\ \emph {et~al.}(2017)\citenamefont
  {Ninarello}, \citenamefont {Berthier},\ and\ \citenamefont
  {Coslovich}}]{ninarello2017models}%
  \BibitemOpen
  \bibfield  {author} {\bibinfo {author} {\bibfnamefont {A.}~\bibnamefont
  {Ninarello}}, \bibinfo {author} {\bibfnamefont {L.}~\bibnamefont
  {Berthier}},\ and\ \bibinfo {author} {\bibfnamefont {D.}~\bibnamefont
  {Coslovich}},\ }\bibfield  {title} {\bibinfo {title} {Models and {A}lgorithms
  for the {N}ext {G}eneration of {G}lass {T}ransition {S}tudies},\ }\href@noop
  {} {\bibfield  {journal} {\bibinfo  {journal} {Phys. Rev. X}\ }\textbf
  {\bibinfo {volume} {7}},\ \bibinfo {pages} {021039} (\bibinfo {year}
  {2017})}\BibitemShut {NoStop}%
\bibitem [{\citenamefont {Pusey}\ \emph {et~al.}(2009)\citenamefont {Pusey},
  \citenamefont {Zaccarelli}, \citenamefont {Valeriani}, \citenamefont {Sanz},
  \citenamefont {Poon},\ and\ \citenamefont {Cates}}]{pusey2009hard}%
  \BibitemOpen
  \bibfield  {author} {\bibinfo {author} {\bibfnamefont {P.}~\bibnamefont
  {Pusey}}, \bibinfo {author} {\bibfnamefont {E.}~\bibnamefont {Zaccarelli}},
  \bibinfo {author} {\bibfnamefont {C.}~\bibnamefont {Valeriani}}, \bibinfo
  {author} {\bibfnamefont {E.}~\bibnamefont {Sanz}}, \bibinfo {author}
  {\bibfnamefont {W.~C.}\ \bibnamefont {Poon}},\ and\ \bibinfo {author}
  {\bibfnamefont {M.~E.}\ \bibnamefont {Cates}},\ }\bibfield  {title} {\bibinfo
  {title} {Hard spheres: crystallization and glass formation},\ }\href@noop {}
  {\bibfield  {journal} {\bibinfo  {journal} {Phil. Trans. R. Soc. A.}\
  }\textbf {\bibinfo {volume} {367}},\ \bibinfo {pages} {4993} (\bibinfo {year}
  {2009})}\BibitemShut {NoStop}%
\bibitem [{\citenamefont {Leocmach}\ \emph {et~al.}(2013)\citenamefont
  {Leocmach}, \citenamefont {Russo},\ and\ \citenamefont
  {Tanaka}}]{leocmach2013importance}%
  \BibitemOpen
  \bibfield  {author} {\bibinfo {author} {\bibfnamefont {M.}~\bibnamefont
  {Leocmach}}, \bibinfo {author} {\bibfnamefont {J.}~\bibnamefont {Russo}},\
  and\ \bibinfo {author} {\bibfnamefont {H.}~\bibnamefont {Tanaka}},\
  }\bibfield  {title} {\bibinfo {title} {Importance of many-body correlations
  in glass transition: {A}n example from polydisperse hard spheres},\
  }\href@noop {} {\bibfield  {journal} {\bibinfo  {journal} {J. Chem. Phys.}\
  }\textbf {\bibinfo {volume} {138}},\ \bibinfo {pages} {12A536} (\bibinfo
  {year} {2013})}\BibitemShut {NoStop}%
\bibitem [{\citenamefont {Li}\ \emph {et~al.}(2016)\citenamefont {Li},
  \citenamefont {Xie}, \citenamefont {Li}, \citenamefont {Qian},\ and\
  \citenamefont {Lu}}]{li2016influence}%
  \BibitemOpen
  \bibfield  {author} {\bibinfo {author} {\bibfnamefont {S.-J.}\ \bibnamefont
  {Li}}, \bibinfo {author} {\bibfnamefont {S.-J.}\ \bibnamefont {Xie}},
  \bibinfo {author} {\bibfnamefont {Y.-C.}\ \bibnamefont {Li}}, \bibinfo
  {author} {\bibfnamefont {H.-J.}\ \bibnamefont {Qian}},\ and\ \bibinfo
  {author} {\bibfnamefont {Z.-Y.}\ \bibnamefont {Lu}},\ }\bibfield  {title}
  {\bibinfo {title} {Influence of molecular-weight polydispersity on the glass
  transition of polymers},\ }\href@noop {} {\bibfield  {journal} {\bibinfo
  {journal} {Phys. Rev. E}\ }\textbf {\bibinfo {volume} {93}},\ \bibinfo
  {pages} {012613} (\bibinfo {year} {2016})}\BibitemShut {NoStop}%
\bibitem [{\citenamefont {Sch{\"o}pe}\ \emph {et~al.}(2007)\citenamefont
  {Sch{\"o}pe}, \citenamefont {Bryant},\ and\ \citenamefont
  {Van~Megen}}]{schope2007effect}%
  \BibitemOpen
  \bibfield  {author} {\bibinfo {author} {\bibfnamefont {H.~J.}\ \bibnamefont
  {Sch{\"o}pe}}, \bibinfo {author} {\bibfnamefont {G.}~\bibnamefont {Bryant}},\
  and\ \bibinfo {author} {\bibfnamefont {W.}~\bibnamefont {Van~Megen}},\
  }\bibfield  {title} {\bibinfo {title} {Effect of polydispersity on the
  crystallization kinetics of suspensions of colloidal hard spheres when
  approaching the glass transition},\ }\href@noop {} {\bibfield  {journal}
  {\bibinfo  {journal} {J. Chem. Phys.}\ }\textbf {\bibinfo {volume} {127}},\
  \bibinfo {pages} {084505} (\bibinfo {year} {2007})}\BibitemShut {NoStop}%
\bibitem [{\citenamefont {Coslovich}\ \emph {et~al.}(2018)\citenamefont
  {Coslovich}, \citenamefont {Ozawa},\ and\ \citenamefont
  {Berthier}}]{coslovich2018local}%
  \BibitemOpen
  \bibfield  {author} {\bibinfo {author} {\bibfnamefont {D.}~\bibnamefont
  {Coslovich}}, \bibinfo {author} {\bibfnamefont {M.}~\bibnamefont {Ozawa}},\
  and\ \bibinfo {author} {\bibfnamefont {L.}~\bibnamefont {Berthier}},\
  }\bibfield  {title} {\bibinfo {title} {Local order and crystallization of
  dense polydisperse hard spheres},\ }\href@noop {} {\bibfield  {journal}
  {\bibinfo  {journal} {J. Phys. Condens. Matter.}\ }\textbf {\bibinfo {volume}
  {30}},\ \bibinfo {pages} {144004} (\bibinfo {year} {2018})}\BibitemShut
  {NoStop}%
\bibitem [{\citenamefont {Baranau}\ and\ \citenamefont
  {Tallarek}(2020)}]{baranau2020relaxation}%
  \BibitemOpen
  \bibfield  {author} {\bibinfo {author} {\bibfnamefont {V.}~\bibnamefont
  {Baranau}}\ and\ \bibinfo {author} {\bibfnamefont {U.}~\bibnamefont
  {Tallarek}},\ }\bibfield  {title} {\bibinfo {title} {Relaxation times,
  jamming densities, and ideal glass transition densities for hard spheres in a
  wide range of polydispersities},\ }\href@noop {} {\bibfield  {journal}
  {\bibinfo  {journal} {AIP Advances}\ }\textbf {\bibinfo {volume} {10}},\
  \bibinfo {pages} {035212} (\bibinfo {year} {2020})}\BibitemShut {NoStop}%
\bibitem [{\citenamefont {Zaccarelli}\ \emph {et~al.}(2015)\citenamefont
  {Zaccarelli}, \citenamefont {Liddle},\ and\ \citenamefont
  {Poon}}]{zaccarelli2015polydispersity}%
  \BibitemOpen
  \bibfield  {author} {\bibinfo {author} {\bibfnamefont {E.}~\bibnamefont
  {Zaccarelli}}, \bibinfo {author} {\bibfnamefont {S.~M.}\ \bibnamefont
  {Liddle}},\ and\ \bibinfo {author} {\bibfnamefont {W.~C.}\ \bibnamefont
  {Poon}},\ }\bibfield  {title} {\bibinfo {title} {On polydispersity and the
  hard sphere glass transition},\ }\href@noop {} {\bibfield  {journal}
  {\bibinfo  {journal} {Soft Matter}\ }\textbf {\bibinfo {volume} {11}},\
  \bibinfo {pages} {324} (\bibinfo {year} {2015})}\BibitemShut {NoStop}%
\bibitem [{\citenamefont {Pihlajamaa}\ \emph {et~al.}(2023)\citenamefont
  {Pihlajamaa}, \citenamefont {Laudicina},\ and\ \citenamefont
  {Janssen}}]{pihlajamaa2023polydispersity}%
  \BibitemOpen
  \bibfield  {author} {\bibinfo {author} {\bibfnamefont {I.}~\bibnamefont
  {Pihlajamaa}}, \bibinfo {author} {\bibfnamefont {C.~C.~L.}\ \bibnamefont
  {Laudicina}},\ and\ \bibinfo {author} {\bibfnamefont {L.~M.~C.}\ \bibnamefont
  {Janssen}},\ }\bibfield  {title} {\bibinfo {title} {Polydispersity modifies
  relaxation mechanisms in glassy liquids},\ }\href@noop {} {\bibfield
  {journal} {\bibinfo  {journal} {arXiv preprint arXiv:2302.09549}\ } (\bibinfo
  {year} {2023})}\BibitemShut {NoStop}%
\bibitem [{\citenamefont {Leutheusser}(1984)}]{leutheusser1984dynamical}%
  \BibitemOpen
  \bibfield  {author} {\bibinfo {author} {\bibfnamefont {E.}~\bibnamefont
  {Leutheusser}},\ }\bibfield  {title} {\bibinfo {title} {Dynamical model of
  the liquid-glass transition},\ }\href@noop {} {\bibfield  {journal} {\bibinfo
   {journal} {Phys. Rev. A}\ }\textbf {\bibinfo {volume} {29}},\ \bibinfo
  {pages} {2765} (\bibinfo {year} {1984})}\BibitemShut {NoStop}%
\bibitem [{\citenamefont {Bengtzelius}\ \emph {et~al.}(1984)\citenamefont
  {Bengtzelius}, \citenamefont {G\"{o}tze},\ and\ \citenamefont
  {Sj\"{o}lander}}]{bengtzelius1984dynamics}%
  \BibitemOpen
  \bibfield  {author} {\bibinfo {author} {\bibfnamefont {U.}~\bibnamefont
  {Bengtzelius}}, \bibinfo {author} {\bibfnamefont {W.}~\bibnamefont
  {G\"{o}tze}},\ and\ \bibinfo {author} {\bibfnamefont {A.}~\bibnamefont
  {Sj\"{o}lander}},\ }\bibfield  {title} {\bibinfo {title} {Dynamics of
  supercooled liquids and the glass transition},\ }\href@noop {} {\bibfield
  {journal} {\bibinfo  {journal} {J. Phys. C Solid State Phys.}\ }\textbf
  {\bibinfo {volume} {17}},\ \bibinfo {pages} {5915} (\bibinfo {year}
  {1984})}\BibitemShut {NoStop}%
\bibitem [{\citenamefont {G{\"o}tze}(2009)}]{gotze2009complex}%
  \BibitemOpen
  \bibfield  {author} {\bibinfo {author} {\bibfnamefont {W.}~\bibnamefont
  {G{\"o}tze}},\ }\href@noop {} {\emph {\bibinfo {title} {Complex {D}ynamics of
  {G}lass-{F}orming {L}iquids: A {M}ode-{C}oupling {T}heory}}},\ Vol.\ \bibinfo
  {volume} {143}\ (\bibinfo  {publisher} {Oxford University Press on Demand},\
  \bibinfo {year} {2009})\BibitemShut {NoStop}%
\bibitem [{\citenamefont {G{\"o}tze}(1996)}]{gotze1996bifurcations}%
  \BibitemOpen
  \bibfield  {author} {\bibinfo {author} {\bibfnamefont {W.}~\bibnamefont
  {G{\"o}tze}},\ }\bibfield  {title} {\bibinfo {title} {Bifurcations of an
  interated mapping with retardations},\ }\href@noop {} {\bibfield  {journal}
  {\bibinfo  {journal} {J. Stat. Phys.}\ }\textbf {\bibinfo {volume} {83}},\
  \bibinfo {pages} {1183} (\bibinfo {year} {1996})}\BibitemShut {NoStop}%
\bibitem [{\citenamefont {Li}\ \emph {et~al.}(1992)\citenamefont {Li},
  \citenamefont {Du}, \citenamefont {Chen}, \citenamefont {Cummins},\ and\
  \citenamefont {Tao}}]{li1992testing}%
  \BibitemOpen
  \bibfield  {author} {\bibinfo {author} {\bibfnamefont {G.}~\bibnamefont
  {Li}}, \bibinfo {author} {\bibfnamefont {W.~M.}\ \bibnamefont {Du}}, \bibinfo
  {author} {\bibfnamefont {X.~K.}\ \bibnamefont {Chen}}, \bibinfo {author}
  {\bibfnamefont {H.~Z.}\ \bibnamefont {Cummins}},\ and\ \bibinfo {author}
  {\bibfnamefont {N.~J.}\ \bibnamefont {Tao}},\ }\bibfield  {title} {\bibinfo
  {title} {Testing {M}ode-{C}oupling {p}redictions for $\alpha$ and $\beta$
  {R}elaxation in ca 0.4 k 0.6 (no 3) 1.4 near the {l}iquid-{g}lass transition
  by light scattering},\ }\href@noop {} {\bibfield  {journal} {\bibinfo
  {journal} {Phys. Rev. A}\ }\textbf {\bibinfo {volume} {45}},\ \bibinfo
  {pages} {3867} (\bibinfo {year} {1992})}\BibitemShut {NoStop}%
\bibitem [{\citenamefont {Kob}\ and\ \citenamefont
  {Andersen}(1995)}]{kob1995testing}%
  \BibitemOpen
  \bibfield  {author} {\bibinfo {author} {\bibfnamefont {W.}~\bibnamefont
  {Kob}}\ and\ \bibinfo {author} {\bibfnamefont {H.~C.}\ \bibnamefont
  {Andersen}},\ }\bibfield  {title} {\bibinfo {title} {Testing
  {M}ode-{C}oupling {T}heory for a supercooled binary {L}ennard-{J}ones
  mixture. {II}. {I}ntermediate scattering function and dynamic
  susceptibility},\ }\href@noop {} {\bibfield  {journal} {\bibinfo  {journal}
  {Phys. Rev. E}\ }\textbf {\bibinfo {volume} {52}},\ \bibinfo {pages} {4134}
  (\bibinfo {year} {1995})}\BibitemShut {NoStop}%
\bibitem [{\citenamefont {G{\"o}tze}(1999)}]{gotze1999recent}%
  \BibitemOpen
  \bibfield  {author} {\bibinfo {author} {\bibfnamefont {W.}~\bibnamefont
  {G{\"o}tze}},\ }\bibfield  {title} {\bibinfo {title} {Recent tests of the
  mode-coupling theory for glassy dynamics},\ }\href@noop {} {\bibfield
  {journal} {\bibinfo  {journal} {J. Phys. Condens. Matter}\ }\textbf {\bibinfo
  {volume} {11}},\ \bibinfo {pages} {A1} (\bibinfo {year} {1999})}\BibitemShut
  {NoStop}%
\bibitem [{\citenamefont {Sciortino}\ and\ \citenamefont
  {Kob}(2001)}]{sciortino2001debye}%
  \BibitemOpen
  \bibfield  {author} {\bibinfo {author} {\bibfnamefont {F.}~\bibnamefont
  {Sciortino}}\ and\ \bibinfo {author} {\bibfnamefont {W.}~\bibnamefont
  {Kob}},\ }\bibfield  {title} {\bibinfo {title} {Debye-{W}aller {F}actor of
  {L}iquid {S}ilica: Theory and {S}imulation},\ }\href@noop {} {\bibfield
  {journal} {\bibinfo  {journal} {Phys. Rev. Lett.}\ }\textbf {\bibinfo
  {volume} {86}},\ \bibinfo {pages} {648} (\bibinfo {year} {2001})}\BibitemShut
  {NoStop}%
\bibitem [{\citenamefont {G{\"o}tze}\ and\ \citenamefont
  {Voigtmann}(2003)}]{gotze2003effect}%
  \BibitemOpen
  \bibfield  {author} {\bibinfo {author} {\bibfnamefont {W.}~\bibnamefont
  {G{\"o}tze}}\ and\ \bibinfo {author} {\bibfnamefont {T.}~\bibnamefont
  {Voigtmann}},\ }\bibfield  {title} {\bibinfo {title} {Effect of composition
  changes on the structural relaxation of a binary mixture},\ }\href@noop {}
  {\bibfield  {journal} {\bibinfo  {journal} {Phys. Rev. E}\ }\textbf {\bibinfo
  {volume} {67}},\ \bibinfo {pages} {021502} (\bibinfo {year}
  {2003})}\BibitemShut {NoStop}%
\bibitem [{\citenamefont {Dawson}\ \emph {et~al.}(2000)\citenamefont {Dawson},
  \citenamefont {Foffi}, \citenamefont {Fuchs}, \citenamefont {G{\"o}tze},
  \citenamefont {Sciortino}, \citenamefont {Sperl}, \citenamefont {Tartaglia},
  \citenamefont {Voigtmann},\ and\ \citenamefont
  {Zaccarelli}}]{dawson2000higher}%
  \BibitemOpen
  \bibfield  {author} {\bibinfo {author} {\bibfnamefont {K.}~\bibnamefont
  {Dawson}}, \bibinfo {author} {\bibfnamefont {G.}~\bibnamefont {Foffi}},
  \bibinfo {author} {\bibfnamefont {M.}~\bibnamefont {Fuchs}}, \bibinfo
  {author} {\bibfnamefont {W.}~\bibnamefont {G{\"o}tze}}, \bibinfo {author}
  {\bibfnamefont {F.}~\bibnamefont {Sciortino}}, \bibinfo {author}
  {\bibfnamefont {M.}~\bibnamefont {Sperl}}, \bibinfo {author} {\bibfnamefont
  {P.}~\bibnamefont {Tartaglia}}, \bibinfo {author} {\bibfnamefont
  {T.}~\bibnamefont {Voigtmann}},\ and\ \bibinfo {author} {\bibfnamefont
  {E.}~\bibnamefont {Zaccarelli}},\ }\bibfield  {title} {\bibinfo {title}
  {Higher-order glass-transition singularities in colloidal systems with
  attractive interactions},\ }\href@noop {} {\bibfield  {journal} {\bibinfo
  {journal} {Phys. Rev. E}\ }\textbf {\bibinfo {volume} {63}},\ \bibinfo
  {pages} {011401} (\bibinfo {year} {2000})}\BibitemShut {NoStop}%
\bibitem [{\citenamefont {Pham}\ \emph {et~al.}(2002)\citenamefont {Pham},
  \citenamefont {Puertas}, \citenamefont {Bergenholtz}, \citenamefont
  {Egelhaaf}, \citenamefont {Moussa{\i}d}, \citenamefont {Pusey}, \citenamefont
  {Schofield}, \citenamefont {Cates}, \citenamefont {Fuchs},\ and\
  \citenamefont {Poon}}]{pham2002multiple}%
  \BibitemOpen
  \bibfield  {author} {\bibinfo {author} {\bibfnamefont {K.~N.}\ \bibnamefont
  {Pham}}, \bibinfo {author} {\bibfnamefont {A.~M.}\ \bibnamefont {Puertas}},
  \bibinfo {author} {\bibfnamefont {J.}~\bibnamefont {Bergenholtz}}, \bibinfo
  {author} {\bibfnamefont {S.~U.}\ \bibnamefont {Egelhaaf}}, \bibinfo {author}
  {\bibfnamefont {A.}~\bibnamefont {Moussa{\i}d}}, \bibinfo {author}
  {\bibfnamefont {P.~N.}\ \bibnamefont {Pusey}}, \bibinfo {author}
  {\bibfnamefont {A.~B.}\ \bibnamefont {Schofield}}, \bibinfo {author}
  {\bibfnamefont {M.~E.}\ \bibnamefont {Cates}}, \bibinfo {author}
  {\bibfnamefont {M.}~\bibnamefont {Fuchs}},\ and\ \bibinfo {author}
  {\bibfnamefont {W.~C.}\ \bibnamefont {Poon}},\ }\bibfield  {title} {\bibinfo
  {title} {Multiple glassy states in a simple model system},\ }\href@noop {}
  {\bibfield  {journal} {\bibinfo  {journal} {Science}\ }\textbf {\bibinfo
  {volume} {296}},\ \bibinfo {pages} {104} (\bibinfo {year}
  {2002})}\BibitemShut {NoStop}%
\bibitem [{\citenamefont {Mandal}\ \emph {et~al.}(2017)\citenamefont {Mandal},
  \citenamefont {Lang}, \citenamefont {Bo{\c{t}}an},\ and\ \citenamefont
  {Franosch}}]{mandal2017nonergodicity}%
  \BibitemOpen
  \bibfield  {author} {\bibinfo {author} {\bibfnamefont {S.}~\bibnamefont
  {Mandal}}, \bibinfo {author} {\bibfnamefont {S.}~\bibnamefont {Lang}},
  \bibinfo {author} {\bibfnamefont {V.}~\bibnamefont {Bo{\c{t}}an}},\ and\
  \bibinfo {author} {\bibfnamefont {T.}~\bibnamefont {Franosch}},\ }\bibfield
  {title} {\bibinfo {title} {Nonergodicity parameters of confined hard-sphere
  glasses},\ }\href@noop {} {\bibfield  {journal} {\bibinfo  {journal} {Soft
  Matter}\ }\textbf {\bibinfo {volume} {13}},\ \bibinfo {pages} {6167}
  (\bibinfo {year} {2017})}\BibitemShut {NoStop}%
\bibitem [{\citenamefont {Fabbian}\ \emph {et~al.}(1999)\citenamefont
  {Fabbian}, \citenamefont {G{\"o}tze}, \citenamefont {Sciortino},
  \citenamefont {Tartaglia},\ and\ \citenamefont {Thiery}}]{fabbian1999ideal}%
  \BibitemOpen
  \bibfield  {author} {\bibinfo {author} {\bibfnamefont {L.}~\bibnamefont
  {Fabbian}}, \bibinfo {author} {\bibfnamefont {W.}~\bibnamefont {G{\"o}tze}},
  \bibinfo {author} {\bibfnamefont {F.}~\bibnamefont {Sciortino}}, \bibinfo
  {author} {\bibfnamefont {P.}~\bibnamefont {Tartaglia}},\ and\ \bibinfo
  {author} {\bibfnamefont {F.}~\bibnamefont {Thiery}},\ }\bibfield  {title}
  {\bibinfo {title} {Ideal glass-glass transitions and logarithmic decay of
  correlations in a simple system},\ }\href@noop {} {\bibfield  {journal}
  {\bibinfo  {journal} {Phys. Rev. E}\ }\textbf {\bibinfo {volume} {59}},\
  \bibinfo {pages} {R1347} (\bibinfo {year} {1999})}\BibitemShut {NoStop}%
\bibitem [{\citenamefont {Voigtmann}(2011)}]{voigtmann2011multiple}%
  \BibitemOpen
  \bibfield  {author} {\bibinfo {author} {\bibfnamefont {T.}~\bibnamefont
  {Voigtmann}},\ }\bibfield  {title} {\bibinfo {title} {Multiple glasses in
  asymmetric binary hard spheres},\ }\href@noop {} {\bibfield  {journal}
  {\bibinfo  {journal} {EPL}\ }\textbf {\bibinfo {volume} {96}},\ \bibinfo
  {pages} {36006} (\bibinfo {year} {2011})}\BibitemShut {NoStop}%
\bibitem [{\citenamefont {Thakur}\ and\ \citenamefont
  {Bosse}(1991)}]{thakur1991glass}%
  \BibitemOpen
  \bibfield  {author} {\bibinfo {author} {\bibfnamefont {J.}~\bibnamefont
  {Thakur}}\ and\ \bibinfo {author} {\bibfnamefont {J.}~\bibnamefont {Bosse}},\
  }\bibfield  {title} {\bibinfo {title} {Glass transition of two-component
  liquids. {II}. the {L}amb-{M}{\"o}ssbauer factors},\ }\href@noop {}
  {\bibfield  {journal} {\bibinfo  {journal} {Phys. Rev. A}\ }\textbf {\bibinfo
  {volume} {43}},\ \bibinfo {pages} {4388} (\bibinfo {year}
  {1991})}\BibitemShut {NoStop}%
\bibitem [{\citenamefont {Biroli}\ \emph {et~al.}(2006)\citenamefont {Biroli},
  \citenamefont {Bouchaud}, \citenamefont {Miyazaki},\ and\ \citenamefont
  {Reichman}}]{biroli2006inhomogeneous}%
  \BibitemOpen
  \bibfield  {author} {\bibinfo {author} {\bibfnamefont {G.}~\bibnamefont
  {Biroli}}, \bibinfo {author} {\bibfnamefont {J.-P.}\ \bibnamefont
  {Bouchaud}}, \bibinfo {author} {\bibfnamefont {K.}~\bibnamefont {Miyazaki}},\
  and\ \bibinfo {author} {\bibfnamefont {D.~R.}\ \bibnamefont {Reichman}},\
  }\bibfield  {title} {\bibinfo {title} {Inhomogeneous {M}ode-{C}oupling
  {T}heory and {G}rowing {D}ynamic {L}ength in {S}upercooled {L}iquids},\
  }\href@noop {} {\bibfield  {journal} {\bibinfo  {journal} {Phys. Rev. Lett.}\
  }\textbf {\bibinfo {volume} {97}},\ \bibinfo {pages} {195701} (\bibinfo
  {year} {2006})}\BibitemShut {NoStop}%
\bibitem [{\citenamefont {Flenner}\ and\ \citenamefont
  {Szamel}(2005)}]{flenner2005relaxation}%
  \BibitemOpen
  \bibfield  {author} {\bibinfo {author} {\bibfnamefont {E.}~\bibnamefont
  {Flenner}}\ and\ \bibinfo {author} {\bibfnamefont {G.}~\bibnamefont
  {Szamel}},\ }\bibfield  {title} {\bibinfo {title} {Relaxation in a glassy
  binary mixture: {M}ode-coupling-like power laws, dynamic heterogeneity, and a
  new non-{G}aussian parameter},\ }\href@noop {} {\bibfield  {journal}
  {\bibinfo  {journal} {Phys. Rev. E}\ }\textbf {\bibinfo {volume} {72}},\
  \bibinfo {pages} {011205} (\bibinfo {year} {2005})}\BibitemShut {NoStop}%
\bibitem [{\citenamefont {Voigtmann}(2003)}]{voigtmann2003mode}%
  \BibitemOpen
  \bibfield  {author} {\bibinfo {author} {\bibfnamefont {T.}~\bibnamefont
  {Voigtmann}},\ }\emph {\bibinfo {title} {Mode-{C}oupling {T}heory of the
  {G}lass {T}ransition in {B}inary {M}ixtures}},\ \href@noop {} {Ph.D.
  thesis},\ \bibinfo  {school} {Technische Universit{\"a}t M{\"u}nchen,
  Universit{\"a}tsbibliothek} (\bibinfo {year} {2003})\BibitemShut {NoStop}%
\bibitem [{\citenamefont {Weysser}\ \emph {et~al.}(2010)\citenamefont
  {Weysser}, \citenamefont {Puertas}, \citenamefont {Fuchs},\ and\
  \citenamefont {Voigtmann}}]{weysser2010structural}%
  \BibitemOpen
  \bibfield  {author} {\bibinfo {author} {\bibfnamefont {F.}~\bibnamefont
  {Weysser}}, \bibinfo {author} {\bibfnamefont {A.~M.}\ \bibnamefont
  {Puertas}}, \bibinfo {author} {\bibfnamefont {M.}~\bibnamefont {Fuchs}},\
  and\ \bibinfo {author} {\bibfnamefont {T.}~\bibnamefont {Voigtmann}},\
  }\bibfield  {title} {\bibinfo {title} {Structural relaxation of polydisperse
  hard spheres: {C}omparison of the {M}ode-{C}oupling {T}heory to a {L}angevin
  dynamics simulation},\ }\href@noop {} {\bibfield  {journal} {\bibinfo
  {journal} {Phys. Rev. E}\ }\textbf {\bibinfo {volume} {82}},\ \bibinfo
  {pages} {011504} (\bibinfo {year} {2010})}\BibitemShut {NoStop}%
\bibitem [{\citenamefont {Pihlajamaa}()}]{MCT_solver}%
  \BibitemOpen
  \bibfield  {author} {\bibinfo {author} {\bibfnamefont {I.}~\bibnamefont
  {Pihlajamaa}},\ }\href
  {\href{https://ilianpihlajamaa.github.io/ModeCouplingTheory.jl/dev/index.html}}
  {\bibinfo {title} {Mode{C}oupling{T}heory.jl v.0.6.2: Generic {S}olver for
  {M}ode-{C}oupling like {I}ntegro-{D}ifferential {E}quations}}\BibitemShut
  {NoStop}%
\bibitem [{\citenamefont {Abraham}\ \emph {et~al.}(2008)\citenamefont
  {Abraham}, \citenamefont {Bhattacharrya},\ and\ \citenamefont
  {Bagchi}}]{abraham2008energy}%
  \BibitemOpen
  \bibfield  {author} {\bibinfo {author} {\bibfnamefont {S.~E.}\ \bibnamefont
  {Abraham}}, \bibinfo {author} {\bibfnamefont {S.~M.}\ \bibnamefont
  {Bhattacharrya}},\ and\ \bibinfo {author} {\bibfnamefont {B.}~\bibnamefont
  {Bagchi}},\ }\bibfield  {title} {\bibinfo {title} {Energy {L}andscape,
  {A}ntiplasticization, and {P}olydispersity {I}nduced {C}rossover of
  {H}eterogeneity in {S}upercooled {P}olydisperse {L}iquids},\ }\href@noop {}
  {\bibfield  {journal} {\bibinfo  {journal} {Phys. Rev. Lett.}\ }\textbf
  {\bibinfo {volume} {100}},\ \bibinfo {pages} {167801} (\bibinfo {year}
  {2008})}\BibitemShut {NoStop}%
\bibitem [{\citenamefont {Mallamace}\ \emph {et~al.}(2010)\citenamefont
  {Mallamace}, \citenamefont {Branca}, \citenamefont {Corsaro}, \citenamefont
  {Leone}, \citenamefont {Spooren}, \citenamefont {Chen},\ and\ \citenamefont
  {Stanley}}]{mallamace2010transport}%
  \BibitemOpen
  \bibfield  {author} {\bibinfo {author} {\bibfnamefont {F.}~\bibnamefont
  {Mallamace}}, \bibinfo {author} {\bibfnamefont {C.}~\bibnamefont {Branca}},
  \bibinfo {author} {\bibfnamefont {C.}~\bibnamefont {Corsaro}}, \bibinfo
  {author} {\bibfnamefont {N.}~\bibnamefont {Leone}}, \bibinfo {author}
  {\bibfnamefont {J.}~\bibnamefont {Spooren}}, \bibinfo {author} {\bibfnamefont
  {S.-H.}\ \bibnamefont {Chen}},\ and\ \bibinfo {author} {\bibfnamefont
  {H.~E.}\ \bibnamefont {Stanley}},\ }\bibfield  {title} {\bibinfo {title}
  {Transport properties of glass-forming liquids suggest that dynamic crossover
  temperature is as important as the glass transition temperature},\
  }\href@noop {} {\bibfield  {journal} {\bibinfo  {journal} {Proc. Natl. Acad.
  Sci.}\ }\textbf {\bibinfo {volume} {107}},\ \bibinfo {pages} {22457}
  (\bibinfo {year} {2010})}\BibitemShut {NoStop}%
\bibitem [{\citenamefont {Berthier}\ and\ \citenamefont
  {Biroli}(2011)}]{berthier2011theoretical}%
  \BibitemOpen
  \bibfield  {author} {\bibinfo {author} {\bibfnamefont {L.}~\bibnamefont
  {Berthier}}\ and\ \bibinfo {author} {\bibfnamefont {G.}~\bibnamefont
  {Biroli}},\ }\bibfield  {title} {\bibinfo {title} {Theoretical perspective on
  the glass transition and amorphous materials},\ }\href@noop {} {\bibfield
  {journal} {\bibinfo  {journal} {Rev. Mod. Phys.}\ }\textbf {\bibinfo {volume}
  {83}},\ \bibinfo {pages} {587} (\bibinfo {year} {2011})}\BibitemShut
  {NoStop}%
\bibitem [{\citenamefont {Vogel}\ and\ \citenamefont
  {Glotzer}(2004)}]{vogel2004spatially}%
  \BibitemOpen
  \bibfield  {author} {\bibinfo {author} {\bibfnamefont {M.}~\bibnamefont
  {Vogel}}\ and\ \bibinfo {author} {\bibfnamefont {S.~C.}\ \bibnamefont
  {Glotzer}},\ }\bibfield  {title} {\bibinfo {title} {Spatially {H}eterogeneous
  {D}ynamics and {D}ynamic {F}acilitation in a {M}odel of {V}iscous {S}ilica},\
  }\href@noop {} {\bibfield  {journal} {\bibinfo  {journal} {Phys. Rev. Lett.}\
  }\textbf {\bibinfo {volume} {92}},\ \bibinfo {pages} {255901} (\bibinfo
  {year} {2004})}\BibitemShut {NoStop}%
\bibitem [{\citenamefont {Bergroth}\ \emph {et~al.}(2005)\citenamefont
  {Bergroth}, \citenamefont {Vogel},\ and\ \citenamefont
  {Glotzer}}]{bergroth2005examination}%
  \BibitemOpen
  \bibfield  {author} {\bibinfo {author} {\bibfnamefont {M.~N.~J.}\
  \bibnamefont {Bergroth}}, \bibinfo {author} {\bibfnamefont {M.}~\bibnamefont
  {Vogel}},\ and\ \bibinfo {author} {\bibfnamefont {S.~C.}\ \bibnamefont
  {Glotzer}},\ }\bibfield  {title} {\bibinfo {title} {Examination of {D}ynamic
  {F}acilitation in {M}olecular {D}ynamics {S}imulations of {G}lass-{F}orming
  {L}iquids},\ }\href@noop {} {\bibfield  {journal} {\bibinfo  {journal} {J.
  Phys. Chem. B}\ }\textbf {\bibinfo {volume} {109}},\ \bibinfo {pages} {6748}
  (\bibinfo {year} {2005})}\BibitemShut {NoStop}%
\bibitem [{\citenamefont {Candelier}\ \emph {et~al.}(2010)\citenamefont
  {Candelier}, \citenamefont {Widmer-Cooper}, \citenamefont {Kummerfeld},
  \citenamefont {Dauchot}, \citenamefont {Biroli}, \citenamefont {Harrowell},\
  and\ \citenamefont {Reichman}}]{candelier2010spatiotemporal}%
  \BibitemOpen
  \bibfield  {author} {\bibinfo {author} {\bibfnamefont {R.}~\bibnamefont
  {Candelier}}, \bibinfo {author} {\bibfnamefont {A.}~\bibnamefont
  {Widmer-Cooper}}, \bibinfo {author} {\bibfnamefont {J.~K.}\ \bibnamefont
  {Kummerfeld}}, \bibinfo {author} {\bibfnamefont {O.}~\bibnamefont {Dauchot}},
  \bibinfo {author} {\bibfnamefont {G.}~\bibnamefont {Biroli}}, \bibinfo
  {author} {\bibfnamefont {P.}~\bibnamefont {Harrowell}},\ and\ \bibinfo
  {author} {\bibfnamefont {D.~R.}\ \bibnamefont {Reichman}},\ }\bibfield
  {title} {\bibinfo {title} {Spatiotemporal hierarchy of relaxation events,
  dynamical heterogeneities, and structural reorganization in a supercooled
  liquid},\ }\href@noop {} {\bibfield  {journal} {\bibinfo  {journal} {Phys.
  Rev. Lett.}\ }\textbf {\bibinfo {volume} {105}},\ \bibinfo {pages} {135702}
  (\bibinfo {year} {2010})}\BibitemShut {NoStop}%
\bibitem [{\citenamefont {Guiselin}\ \emph {et~al.}(2022)\citenamefont
  {Guiselin}, \citenamefont {Scalliet},\ and\ \citenamefont
  {Berthier}}]{guiselin2022microscopic}%
  \BibitemOpen
  \bibfield  {author} {\bibinfo {author} {\bibfnamefont {B.}~\bibnamefont
  {Guiselin}}, \bibinfo {author} {\bibfnamefont {C.}~\bibnamefont {Scalliet}},\
  and\ \bibinfo {author} {\bibfnamefont {L.}~\bibnamefont {Berthier}},\
  }\bibfield  {title} {\bibinfo {title} {Microscopic origin of excess wings in
  relaxation spectra of supercooled liquids},\ }\href@noop {} {\bibfield
  {journal} {\bibinfo  {journal} {Nat. Phys.}\ }\textbf {\bibinfo {volume}
  {18}},\ \bibinfo {pages} {468} (\bibinfo {year} {2022})}\BibitemShut
  {NoStop}%
\bibitem [{\citenamefont {Scalliet}\ \emph {et~al.}(2022)\citenamefont
  {Scalliet}, \citenamefont {Guiselin},\ and\ \citenamefont
  {Berthier}}]{scalliet2022thirty}%
  \BibitemOpen
  \bibfield  {author} {\bibinfo {author} {\bibfnamefont {C.}~\bibnamefont
  {Scalliet}}, \bibinfo {author} {\bibfnamefont {B.}~\bibnamefont {Guiselin}},\
  and\ \bibinfo {author} {\bibfnamefont {L.}~\bibnamefont {Berthier}},\
  }\bibfield  {title} {\bibinfo {title} {Thirty {M}illiseconds in the {L}ife of
  a {S}upercooled {L}iquid},\ }\href@noop {} {\bibfield  {journal} {\bibinfo
  {journal} {Phys. Rev. X}\ }\textbf {\bibinfo {volume} {12}},\ \bibinfo
  {pages} {041028} (\bibinfo {year} {2022})}\BibitemShut {NoStop}%
\bibitem [{\citenamefont {Hansen}\ and\ \citenamefont
  {McDonald}(2013)}]{hansen2013theory}%
  \BibitemOpen
  \bibfield  {author} {\bibinfo {author} {\bibfnamefont {J.-P.}\ \bibnamefont
  {Hansen}}\ and\ \bibinfo {author} {\bibfnamefont {I.~R.}\ \bibnamefont
  {McDonald}},\ }\href@noop {} {\emph {\bibinfo {title} {Theory of {S}imple
  {L}iquids: with {A}pplications to {S}oft {M}atter}}}\ (\bibinfo  {publisher}
  {Academic press},\ \bibinfo {year} {2013})\BibitemShut {NoStop}%
\bibitem [{\citenamefont {Franosch}\ \emph {et~al.}(1997)\citenamefont
  {Franosch}, \citenamefont {Fuchs}, \citenamefont {G{\"o}tze}, \citenamefont
  {Mayr},\ and\ \citenamefont {Singh}}]{franosch1997asymptotic}%
  \BibitemOpen
  \bibfield  {author} {\bibinfo {author} {\bibfnamefont {T.}~\bibnamefont
  {Franosch}}, \bibinfo {author} {\bibfnamefont {M.}~\bibnamefont {Fuchs}},
  \bibinfo {author} {\bibfnamefont {W.}~\bibnamefont {G{\"o}tze}}, \bibinfo
  {author} {\bibfnamefont {M.~R.}\ \bibnamefont {Mayr}},\ and\ \bibinfo
  {author} {\bibfnamefont {A.}~\bibnamefont {Singh}},\ }\bibfield  {title}
  {\bibinfo {title} {Asymptotic laws and preasymptotic correction formulas for
  the relaxation near glass-transition singularities},\ }\href@noop {}
  {\bibfield  {journal} {\bibinfo  {journal} {Phys. Rev. E}\ }\textbf {\bibinfo
  {volume} {55}},\ \bibinfo {pages} {7153} (\bibinfo {year}
  {1997})}\BibitemShut {NoStop}%
\bibitem [{\citenamefont {v.~Schweidler}(1907)}]{vonSchweidler1907studien}%
  \BibitemOpen
  \bibfield  {author} {\bibinfo {author} {\bibfnamefont {E.~R.}\ \bibnamefont
  {v.~Schweidler}},\ }\bibfield  {title} {\bibinfo {title} {Studien {\"u}ber
  die {A}nomalien im {V}erhalten der {D}ielektrika},\ }\href@noop {} {\bibfield
   {journal} {\bibinfo  {journal} {Ann. Phys.}\ }\textbf {\bibinfo {volume}
  {329}},\ \bibinfo {pages} {711} (\bibinfo {year} {1907})}\BibitemShut
  {NoStop}%
\bibitem [{\citenamefont {Baxter}(1970)}]{baxter1970ornstein}%
  \BibitemOpen
  \bibfield  {author} {\bibinfo {author} {\bibfnamefont {R.}~\bibnamefont
  {Baxter}},\ }\bibfield  {title} {\bibinfo {title} {Ornstein--zernike relation
  and percus--yevick approximation for fluid mixtures},\ }\href@noop {}
  {\bibfield  {journal} {\bibinfo  {journal} {J. Chem. Phys.}\ }\textbf
  {\bibinfo {volume} {52}},\ \bibinfo {pages} {4559} (\bibinfo {year}
  {1970})}\BibitemShut {NoStop}%
\bibitem [{\citenamefont {Voigtmann}\ \emph {et~al.}(2004)\citenamefont
  {Voigtmann}, \citenamefont {Puertas},\ and\ \citenamefont
  {Fuchs}}]{voigtmann2004tagged}%
  \BibitemOpen
  \bibfield  {author} {\bibinfo {author} {\bibfnamefont {T.}~\bibnamefont
  {Voigtmann}}, \bibinfo {author} {\bibfnamefont {A.~M.}\ \bibnamefont
  {Puertas}},\ and\ \bibinfo {author} {\bibfnamefont {M.}~\bibnamefont
  {Fuchs}},\ }\bibfield  {title} {\bibinfo {title} {Tagged-particle dynamics in
  a hard-sphere system: Mode-coupling theory analysis},\ }\href@noop {}
  {\bibfield  {journal} {\bibinfo  {journal} {Phys. Rev. E}\ }\textbf {\bibinfo
  {volume} {70}},\ \bibinfo {pages} {061506} (\bibinfo {year}
  {2004})}\BibitemShut {NoStop}%
\bibitem [{\citenamefont {Frenkel}\ \emph {et~al.}(1986)\citenamefont
  {Frenkel}, \citenamefont {Vos}, \citenamefont {de~Kruif},\ and\ \citenamefont
  {Vrij}}]{frenkel1986structure}%
  \BibitemOpen
  \bibfield  {author} {\bibinfo {author} {\bibfnamefont {D.}~\bibnamefont
  {Frenkel}}, \bibinfo {author} {\bibfnamefont {R.}~\bibnamefont {Vos}},
  \bibinfo {author} {\bibfnamefont {C.~d.}\ \bibnamefont {de~Kruif}},\ and\
  \bibinfo {author} {\bibfnamefont {A.}~\bibnamefont {Vrij}},\ }\bibfield
  {title} {\bibinfo {title} {Structure factors of polydisperse systems of hard
  spheres: A comparison of monte carlo simulations and percus--yevick theory},\
  }\href@noop {} {\bibfield  {journal} {\bibinfo  {journal} {J. Chem. Phys.}\
  }\textbf {\bibinfo {volume} {84}},\ \bibinfo {pages} {4625} (\bibinfo {year}
  {1986})}\BibitemShut {NoStop}%
\bibitem [{\citenamefont {Fuchs}\ \emph {et~al.}(1991)\citenamefont {Fuchs},
  \citenamefont {G\"{o}tze}, \citenamefont {Hofacker},\ and\ \citenamefont
  {Latz}}]{fuchs1991comments}%
  \BibitemOpen
  \bibfield  {author} {\bibinfo {author} {\bibfnamefont {M.}~\bibnamefont
  {Fuchs}}, \bibinfo {author} {\bibfnamefont {W.}~\bibnamefont {G\"{o}tze}},
  \bibinfo {author} {\bibfnamefont {I.}~\bibnamefont {Hofacker}},\ and\
  \bibinfo {author} {\bibfnamefont {A.}~\bibnamefont {Latz}},\ }\bibfield
  {title} {\bibinfo {title} {Comments on the alpha-peak shapes for relaxation
  in supercooled liquids},\ }\href@noop {} {\bibfield  {journal} {\bibinfo
  {journal} {J. Phys. Condens. Matter}\ }\textbf {\bibinfo {volume} {3}},\
  \bibinfo {pages} {5047} (\bibinfo {year} {1991})}\BibitemShut {NoStop}%
\bibitem [{\citenamefont {G{\"o}tze}(1991)}]{gotze1991liquids}%
  \BibitemOpen
  \bibfield  {author} {\bibinfo {author} {\bibfnamefont {W.}~\bibnamefont
  {G{\"o}tze}},\ }\href@noop {} {\bibinfo {title} {Liquids, {F}reezing and the
  {G}lass {T}ransition}} (\bibinfo {year} {1991})\BibitemShut {NoStop}%
\bibitem [{\citenamefont {Foffi}\ \emph {et~al.}(2003)\citenamefont {Foffi},
  \citenamefont {G{\"o}tze}, \citenamefont {Sciortino}, \citenamefont
  {Tartaglia},\ and\ \citenamefont {Voigtmann}}]{foffi2003mixing}%
  \BibitemOpen
  \bibfield  {author} {\bibinfo {author} {\bibfnamefont {G.}~\bibnamefont
  {Foffi}}, \bibinfo {author} {\bibfnamefont {W.}~\bibnamefont {G{\"o}tze}},
  \bibinfo {author} {\bibfnamefont {F.}~\bibnamefont {Sciortino}}, \bibinfo
  {author} {\bibfnamefont {P.}~\bibnamefont {Tartaglia}},\ and\ \bibinfo
  {author} {\bibfnamefont {T.}~\bibnamefont {Voigtmann}},\ }\bibfield  {title}
  {\bibinfo {title} {Mixing {E}ffects for the {S}tructural {R}elaxation in
  {B}inary {H}ard-{S}phere {L}iquids},\ }\href@noop {} {\bibfield  {journal}
  {\bibinfo  {journal} {Phys. Rev. Lett.}\ }\textbf {\bibinfo {volume} {91}},\
  \bibinfo {pages} {085701} (\bibinfo {year} {2003})}\BibitemShut {NoStop}%
\bibitem [{\citenamefont {Sciortino}\ and\ \citenamefont
  {Tartaglia}(2005)}]{sciortino2005glassy}%
  \BibitemOpen
  \bibfield  {author} {\bibinfo {author} {\bibfnamefont {F.}~\bibnamefont
  {Sciortino}}\ and\ \bibinfo {author} {\bibfnamefont {P.}~\bibnamefont
  {Tartaglia}},\ }\bibfield  {title} {\bibinfo {title} {Glassy {C}olloidal
  {S}ystems},\ }\href@noop {} {\bibfield  {journal} {\bibinfo  {journal}
  {Advances in Physics}\ }\textbf {\bibinfo {volume} {54}},\ \bibinfo {pages}
  {471} (\bibinfo {year} {2005})}\BibitemShut {NoStop}%
\bibitem [{\citenamefont {Asakura}\ and\ \citenamefont
  {Oosawa}(1958)}]{asakura1958interaction}%
  \BibitemOpen
  \bibfield  {author} {\bibinfo {author} {\bibfnamefont {S.}~\bibnamefont
  {Asakura}}\ and\ \bibinfo {author} {\bibfnamefont {F.}~\bibnamefont
  {Oosawa}},\ }\bibfield  {title} {\bibinfo {title} {Interaction between
  particles suspended in solutions of macromolecules},\ }\href@noop {}
  {\bibfield  {journal} {\bibinfo  {journal} {J. Polym. Sci.}\ }\textbf
  {\bibinfo {volume} {33}},\ \bibinfo {pages} {183} (\bibinfo {year}
  {1958})}\BibitemShut {NoStop}%
\bibitem [{\citenamefont {Bergenholtz}\ and\ \citenamefont
  {Fuchs}(1999)}]{bergenholtz1999nonergodicity}%
  \BibitemOpen
  \bibfield  {author} {\bibinfo {author} {\bibfnamefont {J.}~\bibnamefont
  {Bergenholtz}}\ and\ \bibinfo {author} {\bibfnamefont {M.}~\bibnamefont
  {Fuchs}},\ }\bibfield  {title} {\bibinfo {title} {Nonergodicity transitions
  in colloidal suspensions with attractive interactions},\ }\href@noop {}
  {\bibfield  {journal} {\bibinfo  {journal} {Phys. Rev. E}\ }\textbf {\bibinfo
  {volume} {59}},\ \bibinfo {pages} {5706} (\bibinfo {year}
  {1999})}\BibitemShut {NoStop}%
\bibitem [{\citenamefont {Heckendorf}\ \emph {et~al.}(2017)\citenamefont
  {Heckendorf}, \citenamefont {Mutch}, \citenamefont {Egelhaaf},\ and\
  \citenamefont {Laurati}}]{heckendorf2017size}%
  \BibitemOpen
  \bibfield  {author} {\bibinfo {author} {\bibfnamefont {D.}~\bibnamefont
  {Heckendorf}}, \bibinfo {author} {\bibfnamefont {K.~J.}\ \bibnamefont
  {Mutch}}, \bibinfo {author} {\bibfnamefont {S.~U.}\ \bibnamefont
  {Egelhaaf}},\ and\ \bibinfo {author} {\bibfnamefont {M.}~\bibnamefont
  {Laurati}},\ }\bibfield  {title} {\bibinfo {title} {Size-{D}ependent
  {L}ocalization in {P}olydisperse {C}olloidal {G}lasses},\ }\href@noop {}
  {\bibfield  {journal} {\bibinfo  {journal} {Phys. Rev. Lett.}\ }\textbf
  {\bibinfo {volume} {119}},\ \bibinfo {pages} {048003} (\bibinfo {year}
  {2017})}\BibitemShut {NoStop}%
\bibitem [{\citenamefont {Debenedetti}\ and\ \citenamefont
  {Stillinger}(2001)}]{debenedetti2001supercooled}%
  \BibitemOpen
  \bibfield  {author} {\bibinfo {author} {\bibfnamefont {P.~G.}\ \bibnamefont
  {Debenedetti}}\ and\ \bibinfo {author} {\bibfnamefont {F.~H.}\ \bibnamefont
  {Stillinger}},\ }\bibfield  {title} {\bibinfo {title} {Supercooled liquids
  and the glass transition},\ }\href@noop {} {\bibfield  {journal} {\bibinfo
  {journal} {Nature}\ }\textbf {\bibinfo {volume} {410}},\ \bibinfo {pages}
  {259} (\bibinfo {year} {2001})}\BibitemShut {NoStop}%
\bibitem [{\citenamefont {Phillips}(1996)}]{phillips1996stretched}%
  \BibitemOpen
  \bibfield  {author} {\bibinfo {author} {\bibfnamefont {J.~C.}\ \bibnamefont
  {Phillips}},\ }\bibfield  {title} {\bibinfo {title} {Stretched exponential
  relaxation in molecular and electronic glasses},\ }\href@noop {} {\bibfield
  {journal} {\bibinfo  {journal} {Rep. Prog. Phys.}\ }\textbf {\bibinfo
  {volume} {59}},\ \bibinfo {pages} {1133} (\bibinfo {year}
  {1996})}\BibitemShut {NoStop}%
\bibitem [{\citenamefont {Fuchs}(1994)}]{fuchs1994kohlrausch}%
  \BibitemOpen
  \bibfield  {author} {\bibinfo {author} {\bibfnamefont {M.}~\bibnamefont
  {Fuchs}},\ }\bibfield  {title} {\bibinfo {title} {The kohlrausch law as a
  limit solution to mode coupling equations},\ }\href@noop {} {\bibfield
  {journal} {\bibinfo  {journal} {J. Non Cryst. Solids}\ }\textbf {\bibinfo
  {volume} {172}},\ \bibinfo {pages} {241} (\bibinfo {year}
  {1994})}\BibitemShut {NoStop}%
\bibitem [{\citenamefont {Blochowicz}\ \emph {et~al.}(2003)\citenamefont
  {Blochowicz}, \citenamefont {Tschirwitz}, \citenamefont {Benkhof},\ and\
  \citenamefont {R{\"o}ssler}}]{blochowicz2003susceptibility}%
  \BibitemOpen
  \bibfield  {author} {\bibinfo {author} {\bibfnamefont {T.}~\bibnamefont
  {Blochowicz}}, \bibinfo {author} {\bibfnamefont {C.}~\bibnamefont
  {Tschirwitz}}, \bibinfo {author} {\bibfnamefont {S.}~\bibnamefont
  {Benkhof}},\ and\ \bibinfo {author} {\bibfnamefont {E.}~\bibnamefont
  {R{\"o}ssler}},\ }\bibfield  {title} {\bibinfo {title} {Susceptibility
  functions for slow relaxation processes in supercooled liquids and the search
  for universal relaxation patterns},\ }\href@noop {} {\bibfield  {journal}
  {\bibinfo  {journal} {J. Chem. Phys.}\ }\textbf {\bibinfo {volume} {118}},\
  \bibinfo {pages} {7544} (\bibinfo {year} {2003})}\BibitemShut {NoStop}%
\bibitem [{\citenamefont {Szamel}\ and\ \citenamefont
  {Flenner}(2010)}]{szamel2010diverging}%
  \BibitemOpen
  \bibfield  {author} {\bibinfo {author} {\bibfnamefont {G.}~\bibnamefont
  {Szamel}}\ and\ \bibinfo {author} {\bibfnamefont {E.}~\bibnamefont
  {Flenner}},\ }\bibfield  {title} {\bibinfo {title} {Diverging length scale of
  the inhomogeneous mode-coupling theory: {A} numerical investigation},\
  }\href@noop {} {\bibfield  {journal} {\bibinfo  {journal} {Phys. Rev. E}\
  }\textbf {\bibinfo {volume} {81}},\ \bibinfo {pages} {031507} (\bibinfo
  {year} {2010})}\BibitemShut {NoStop}%
\bibitem [{\citenamefont {H{\"o}fling}\ \emph {et~al.}(2006)\citenamefont
  {H{\"o}fling}, \citenamefont {Franosch},\ and\ \citenamefont
  {Frey}}]{hofling2006localization}%
  \BibitemOpen
  \bibfield  {author} {\bibinfo {author} {\bibfnamefont {F.}~\bibnamefont
  {H{\"o}fling}}, \bibinfo {author} {\bibfnamefont {T.}~\bibnamefont
  {Franosch}},\ and\ \bibinfo {author} {\bibfnamefont {E.}~\bibnamefont
  {Frey}},\ }\bibfield  {title} {\bibinfo {title} {Localization transition of
  the three-dimensional lorentz model and continuum percolation},\ }\href@noop
  {} {\bibfield  {journal} {\bibinfo  {journal} {Phys. Rev. Lett.}\ }\textbf
  {\bibinfo {volume} {96}},\ \bibinfo {pages} {165901} (\bibinfo {year}
  {2006})}\BibitemShut {NoStop}%
\end{thebibliography}%

\end{document}